\documentclass[11pt,epsf]{article}

\usepackage{amsmath}
\usepackage{amssymb}
\usepackage{amsthm}
\usepackage{epsfig}
\usepackage[dvipsnames]{xcolor}

\newcommand{\prt}{\partial}

\def\ni{{\noindent}}

\def\be{\begin{equation}}
\def\ee{\end{equation}}
\numberwithin{equation}{section}

\theoremstyle{definition}

\newcommand{\comment}[1]{}

\begin{document}
\bibliographystyle{plain}

\title{{\Large\bf On the Whitham system for the (2+1)-dimensional nonlinear Schr\"odinger equation}}
\author{Mark~J.~Ablowitz$^1$, Justin~T.~Cole$^2$, Igor~Rumanov$^1$\footnote{corresponding author; e-mail: igor.rumanov@colorado.edu} \\
{\small $^1$Department of Applied Mathematics, University of Colorado at Boulder} \\    
{\small $^2$Department of Mathematics, University of Colorado at Colorado Springs}}

\maketitle

\bigskip

\begin{abstract}
We derive the Whitham modulation equations for the nonlinear Schr\"odinger equation in the plane (2d NLS) with small dispersion. The modulation equations are derived in terms of both physical and Riemann variables; the latter yields equations of hydrodynamic type. The complete 2d NLS Whitham system consists of six dynamical equations in evolutionary form and two constraints. As an application, we determine the linear stability of one-dimensional traveling waves. In both the elliptic and hyperbolic case, the traveling waves are found to be unstable. This result is consistent with all previous investigations of such stability by other methods. These results are supported by direct numerical calculations. 

\end{abstract}

\section{Introduction}  

The (2+1)-dimensional nonlinear Schr\"odinger (2d NLS) equation is

\be
i\epsilon\prt_t\Psi + \epsilon^2(\prt_{xx}+s\prt_{yy})\Psi - |\Psi|^2\Psi = 0, ~~\Psi(x,y,t=0) =\Psi_{in}(x,y), ~~|x|< \infty, |y|< \infty      \label{eq:2dNLS}   
\ee
where $s=\pm1$ is a sign distinguishing the elliptic ($s=1$) from hyperbolic ($s=-1$) 2d NLS equation. In the elliptic case, this is the defocusing 2dNLS equation. The applicability of the 2d NLS equation is well-known; for example, deep water waves with surface tension~\cite{AbSeg79, AbSeg81, SulSul99}, Bose-Einstein condensate (BEC) wavefunctions~\cite{PitStrBEC, Fr2010} in appropriate geometries with the third coordinate axis being the axis of symmetry, plasma physics~\cite{Ber98, SulSul99}, nonlinear optics~\cite{SulSul99, WJF-NatPhys07}, magnetics, and many others.

Here Whitham modulation theory~\cite{Whitham74} is used to study the 2d NLS equation with small dispersion. This is a natural extension of earlier work beginning with the cylindrical KdV reduction of Kadomtsev-Petviashvili (KP) equation and reduction of (2+1)-dimensional Benjamin-Ono (BO) equation~\cite{ADM}. This was followed by the derivation of full Whitham systems for the KP, (2+1)d BO and a modified KP equation~\cite{ABW-KP, ABW-BO, ABR}. The Whitham system was subsequently derived~\cite{rNLS} for the radial NLS (rNLS) reduction of the elliptic 2d NLS equation, which is relevant e.g.~in experiments with laser-induced ring-shaped humps of BEC density~\cite{HoeferEtAl06}, and in fact in many other media where 2d NLS is an effective model. Curved propagating DSW fronts are often observed, see e.g.~\cite{WJF-NatPhys07, GarTrilloEtAl15, HyperNLS}. In this paper, the complete Whitham system for the 2d NLS equation, both elliptic and hyperbolic, is constructed. Besides applying to small dispersion regimes, as a WKB-type theory it also describes slowly varying  waves with finite dispersion and amplitude, see e.g.~\cite{Ab2011}.

The above developments are important steps toward understanding systems with spatial dimensions higher than one by means of the Whitham multiscale approach. Whitham modulation theory~\cite{Whitham74} for spatially one-dimensional partial differential equations (PDEs) has been extensively studied for already more than five decades. It is reviewed e.g.~in~\cite{ElHoefRev16} which contains the description of main mathematical methods, experimental and numerical results and a comprehensive list of references. Important equations and their typical dispersive shock wave (DSW) solutions analyzed by Whitham theory include the integrable Korteweg-de Vries (KdV)~\cite{Whitham74, GurPit73} and the 1d NLS equation~\cite{ForLee86, Pa87, GK87}. A Riemann-type step problem for the 1d NLS equation leading to the DSW formation was first considered in~\cite{GK87}. The theory of one-dimensional DSWs in the NLS equation was further developed by classifying the possible types of initial discontinuities~\cite{ElGGK95}.

The number of theoretical results available for nonlinear waves in (1+1)-dimensional systems is huge compared with those for higher dimensions. Yet most often the special experimental setups~\cite{TrilloXuEtAl17, JanSprenHoefWu17} are needed to create a physical situation where e.g.~1d NLS nonlinear waves, constant amplitude periodic traveling waves or DSWs are a good approximation. On the other hand, in most physical systems like water waves, BECs, plasmas, magnetics, the nonlinear wave phenomena of higher dimensionality naturally appear which require higher-dimensional theoretical methods to understand them. For example, the radial NLS Whitham system~\cite{rNLS} better describes the experiments of~\cite{HoeferEtAl06} than does the 1d approximation originally used there. Numerous one-dimensional periodic waves are known to be unstable to transverse perturbations; this instability can lead to the formation of various higher-dimensional structures like e.g.~vortices~\cite{PSK95, TheoEtAl03, ChSch18} which can be described only in 2d NLS (or 3d NLS) setting. Thus the significance of developing the more challenging theories and methods for higher dimensions cannot be overestimated.

As indicated above, there are a number of existing results in literature about transverse instabilities of one dimensional nonlinear waves embedded within the 2d NLS equation. The first work dealt with bright one dimensional solitons in focusing NLS~\cite{ZaRub73}. The transverse instability investigation of dark solitons in the defocusing NLS equation was carried out in~\cite{KuzTur88}. The stability of traveling wave solutions was investigated in numerous places using various methods; see e.g.~\cite{CarterThesis, CarSeg03, KartEtAl03, TheCarDecon06, HoefIl12}. Since the theory we develop here is algebraically complex, it is important to find sufficiently simple applications to test it. We consider one such application -- the investigation of linear stability of exact periodic flat traveling wave solutions, which are also a basis for the modulated DSW solutions. Our finding that one-dimensional periodic waves are transversely unstable is consistent with the above instability results. 

\par The outline of the paper is the following. In section 2 we describe the main results of the paper. In section 3 we apply the direct perturbation version of Whitham approach with fast and slow space and time scales and derive the leading order solution. In section 4 we use the Euler hydrodynamic representation of the 2d NLS equation to derive the Whitham modulation system in `physical' variables including mass, momentum, energy and their fluxes. Then, in section 5, together with technical details in appendices B and C, we express the system in terms of four 1d NLS Riemann variables and two additional variables pertaining to the 2d case. Thus we obtain the main equations eq.~(\ref{eq:Wr-eq}), (\ref{eq:q-eq}) and (\ref{eq:p-eq}). In section 6 we present the linear stability analysis for the one-dimensional traveling wave solutions based on our Whitham PDE system and compare with the direct numerics of the 2d NLS equation. Section 7 presents our conclusions and future directions. Appendix A contains some auxiliary formulas from the theory of elliptic functions, see e.g.~\cite{BF71}, in relation with the leading order solution. In Appendix B, an auxiliary intermediate Whitham system in physical variables is derived from the original one. It facilitates the transition to the 1d NLS-type Riemann variables in section 5, with supporting details in Appendix C.

\section{Main results}

In this paper we derive the Whitham equations for slow modulations in the 2d NLS equation with small dispersion, first in physical variables and then in terms of so-called Riemann variables. The latter is a hydrodynamic like system of equations; these equations are the analogs of the Riemann hydrodynamic equations found for the 1d NLS~\cite{ForLee86, Pa87, GK87} and rNLS~\cite{rNLS} Whitham systems. Transforming Whitham equations from physical variables to hydrodynamic Riemann variables is a common feature of the Whitham procedure and is carried out for the 2d NLS equation discussed here. The Riemann-type variables in the 2d NLS Whitham system of this paper are also the analogs of the Riemann invariants of the 1d NLS Whitham system~\cite{ForLee86, Pa87, GK87,HoeferEtAl06} and Riemann-type variables of the rNLS Whitham system~\cite{rNLS}.

The 2d NLS Whitham system given below in the Riemann variables: $\{r_j, j=1,2,3,4\}$, and additional variables $q, p$, is a major result of this paper:
\be
\prt_tr_j + v_j\prt_xr_j + s(qv_j+2p)\prt_yr_j + h_j(r_1,r_2,r_3,r_4,q,p) = 0,   \quad j=1,2,3,4  \label{eq:Wr-eq}    
\ee
here $v_j$ are the well-known 1d NLS Whitham velocities~\cite{GK87, Pa87}, 

{\small$$
v_1 = V - \frac{(r_2-r_1)(r_4-r_1)}{2(r_4-r_1 - (r_4-r_2)E(m)/K(m))},    \qquad   v_2 = V + \frac{(r_2-r_1)(r_3-r_2)}{2(r_3-r_2 - (r_3-r_1)E(m)/K(m))},
$$

\be
v_3 = V - \frac{(r_4-r_3)(r_3-r_2)}{2(r_3-r_2 - (r_4-r_2)E(m)/K(m))},   \qquad   v_4 = V + \frac{(r_4-r_3)(r_4-r_1)}{2(r_4-r_1 - (r_3-r_1)E(m)/K(m))},    \label{eq:vs}
\ee}
where 
\be
V= \frac{r_1+r_2+r_3+r_4}{4},  \qquad  m = \frac{(r_2-r_1)(r_4-r_3)}{(r_3-r_1)(r_4-r_2)},    \label{eq:k2r-eq}    
\ee
and $K(m)$ and $E(m)$ are the complete elliptic integrals of 1st and 2nd kind, respectively. The terms $h_j$ are given by

\be
h_j = sq\left[ (2V-r_j)D_yV - \frac{D_yR_2}{2} + r_j(v_j-V)\prt_xq + 2(v_j-V)\prt_xp \right] + s\left[ Y_jD_yq + Z_jD_yp\right],    \label{eq:hj-eq}
\ee
where

$$
R_2 = \frac{\sum_{j=1}^4r_j^2}{4},  \qquad D_y = \prt_y - q\prt_x, \qquad g \equiv 1+sq^2,        
$$
$$
Y_j = \left(1+\frac{2sq^2}{g}\right)\frac{[(r_j-V)(v_j-V)-R_2+V^2]}{3} + \left(1-\frac{2sq^2}{g}\right)V(r_j-v_j),   \quad  Z_j = 2\left(1-\frac{2sq^2}{g}\right)(r_j-v_j).
$$
The other equations are

\be
\prt_tq + (gV+2sqp)\prt_xq + (\prt_y-q\prt_x)(gV+2sqp) = 0,   \label{eq:q-eq}
\ee


\be
2(\prt_tp + gV\prt_xp + 2sp\prt_yp) - \frac{(\prt_y - q\prt_x)[g(R_2-2V^2)]}{2} = 0.   \label{eq:p-eq}
\ee
Besides, there are two important constraints, 

\be
\prt_xq = \frac{(\prt_y-q\prt_x)k}{k},  \qquad   k^2 = \frac{(r_3-r_1)(r_4-r_2)}{64K^2(m)},    \label{eq:kq-eq}
\ee
    
\be
\prt_xp = \sum_j r_j\frac{\prt_jk}{k}(\prt_y-q\prt_x)r_j - 2V\frac{(\prt_y-q\prt_x)k}{k},   \qquad \prt_j \equiv \frac{\prt}{\prt r_j}.   \label{eq:px-eq}
\ee
Thus, we have six dynamical equations and two constraints (which must be imposed initially) for six dependent variables $r_j, j=1,2,3, 4$, $q$ and $p$.

The above equations while complicated in form are nevertheless a significant simplification of the Whitham equations in physical variables because the time derivatives of each unknown are explicitly given in terms of the six unknowns and their spatial derivatives, plus two constraints to be imposed initially. This evolutionary form of the 2d NLS Whitham system is clearly advantageous e.g.~if one is to solve an initial value problem for the PDEs numerically. Besides, one can compare and see similarities and differences of the above system with e.g.~the 1d NLS Whitham or the KP Whitham system.

\par Our derivation of the Whitham-2dNLS system of eqs.~(\ref{eq:Wr-eq}), (\ref{eq:q-eq}), (\ref{eq:p-eq}) with the additional constraints (\ref{eq:kq-eq}) and (\ref{eq:px-eq}) in some respects parallels that of the well-known 1d NLS Whitham system, while in other respects parallels that of the Whitham systems for the PDEs of KP type~\cite{ABR}. The system and the derivation are, however, algebraically more complicated than previous cases. We believe that the importance of the 2d NLS equations, both elliptic and hyperbolic, justifies the effort to obtain their complete Whitham theory. The derivation given in detail in this paper is a necessary step toward this goal. 

\par In section 3 the modulated periodic solution of the 2dNLS equation is found; it has the same form as that of the 1dNLS equation. The modulated leading order solution for $\Psi=\rho^{1/2}e^{i\Theta}$, $\rho=|\Psi|^2$, can be reconstructed in terms of the variables $r_j, j=1,2,3,4, q, p$ as follows: 

$$
\Psi_0 = \rho_0^{1/2}(\theta; x, y, t; \epsilon)e^{i\Theta_0(\theta; x, y, t; \epsilon)},
$$

\be
\rho_0 = (1+sq^2)\left[\frac{(r_2+r_4-r_1-r_3)^2}{32} - \frac{(r_2-r_1)(r_4-r_3)}{8}\;\text{cn}^2\left(2K(m)(\theta - \theta_*); m\right)\right],   \label{eq:rho0-eq}   
\ee
where the fast phase $\theta$ is determined by the formula

$$
\theta(x,y,t; \epsilon) = \int_0^x \prt_x\theta(z,y,t)dz + \int_0^y \prt_y\theta(0,z,t)dz + \int_0^t \prt_t\theta(0,0,\tau)d\tau,
$$

\be
\epsilon\prt_x\theta = k,  \qquad  \epsilon\prt_y\theta = kq,  \qquad  \epsilon\prt_t\theta = -k[(1+sq^2)V + 2sqp],  \label{eq:theta}  
\ee
and $\theta_*=\theta_*(x,y,t)$ is an order unity phase shift which we will not discuss here. The associated analogs of the hydrodynamic velocity components $u = \epsilon\prt_x\Theta$, $v = \epsilon\prt_y\Theta$, are to leading order
{\small\be
u_0 = \frac{V}{2} + \frac{(1+sq^2)C_0}{2\rho_0},   \qquad  v_0 = qu_0+p,  \qquad C_0 = -\frac{(r_1+r_4-r_2-r_3)(r_1+r_3-r_2-r_4)(r_1+r_2-r_3-r_4)}{128},  \label{eq:u0-eq}     
\ee}
in terms of the introduced variables $r_j$, $q$ and $p$. The total leading order phase of the wave function $\Theta_0$ can be determined e.g.~by the formula

\small{\be
\Theta_0 = -\frac{C_0}{2k}\int_0^{\theta(x,y,t; \epsilon)}\frac{(1+sq^2)dz}{\rho_0(z; x, y, t)} + \frac{\phi(x,y,t)}{\epsilon},   \quad \prt_x\phi = \frac{V}{2},  \quad  \prt_y\phi = \frac{qV+2p}{2},  \quad \prt_t\phi = -\frac{gR_2+4sp(qV+p)}{4},   \label{eq:Phase}
\ee}
where $\phi(x,y,t)$ is restored from its partials above similarly to $\theta$ in eq.~(\ref{eq:theta}) i.e.
\be
\phi = \int_0^x\prt_x\phi(z,y,t)dz + \int_0^y\prt_y\phi(0,z,t)dz + \int_0^t\prt_t\phi(0,0,\tau)d\tau.   \label{eq:phi-eq}
\ee

{\it Remark 1.} We should emphasize, however, that the above solution $\Psi_0$ describes more general periodic traveling structures than its 1d NLS analog. E.g.~some particular cases are the 1d periodic (plane) traveling wave and the radial (ring) periodic traveling wave which is the leading order solution of rNLS~\cite{rNLS}, or more general periodic spatially two-dimensional waves (like spiral waves etc.).
\par {\it Remark 2.} It is simple but instructive to see how the 2dNLS Whitham system yields the one-dimensional result for the oblique propagation in arbitrary direction. Then one should take $q=\text{const.}$ and $p=0$. Immediately one finds from eqs.~(\ref{eq:q-eq}) and (\ref{eq:p-eq}) that $D_yV=0$ and $D_yR_2=0$ must hold. This implies $h_j=0$ in eqs.~(\ref{eq:Wr-eq}). Besides, the constraints (\ref{eq:kq-eq}) and (\ref{eq:px-eq}) give two more linear homogeneous equations for the derivatives $D_yr_j$: $D_yk=0$ and $\sum_j r_j\frac{\prt_jk}{k}D_yr_j=0$. One concludes that all $D_y=\prt_y-q\prt_x$-derivatives must be zero. Then one is left with the four equations (\ref{eq:Wr-eq}) which now reduce to $\prt_tr_j + v_j(\prt_x+sq\prt_y)r_j=0$, which is the well-known 1dNLS Whitham system up to a linear change of the spatial coordinates. 

\par {\it Remark 3.} For the hyperbolic 2d NLS ($s=-1$) case, the density $\rho_0$ in eq.~(\ref{eq:rho0-eq}) turns to zero when $q=\pm1$ and appears to change sign if $|q|$ exceeds one, which is inconsistent with $\rho_0$ being nonnegative by definition. This is related to the fact that our hyperbolic 2d NLS equation is defocusing in the $x$- and focusing in the $y$-direction. The character of its leading order solution changes for $|q|>1$, which is not covered by the above formulas. The full meaning and consequences of this remain to be investigated; but this is postponed to a future publication. Thus, it is safe to say that our hyperbolic 2d NLS Whitham system and leading order solution are valid for $|q|<1$, while no such restriction exists for the elliptic 2d NLS (defocusing in all directions).

\medskip

\par We also analyzed the linear stability of one-dimensional traveling wave solutions of the 2d NLS equation using the Whitham theory developed in this paper. The Whitham approach consists of considering small periodic perturbations to the six Whitham equations displayed above. This is the method used before in e.g.~\cite{ABW-KP, ABW-BO, ABR} to compute the linear stability of traveling waves for (2+1)-dimensional equations of KP-type. As a result, we obtained the two transverse perturbation modes $\omega_{\pm}(m)$, one of them unstable for the elliptic 2d NLS and the other -- for the hyperbolic NLS equation, in the full range of elliptic modulus $m$, $0<m<1$, and the other parameters. Thus our Whitham theory allows us to find out that both elliptic and hyperbolic 2d NLS equations are linearly unstable with respect to small transverse perturbations, and the corresponding instability growth rates. The results are compared with direct numerics, for small transverse perturbation wavenumbers $l$ where Whitham theory is valid; very good agreement is found. They also agree with the previous results of direct perturbation analysis e.g.~in~\cite{CarterThesis, CarSeg03, KuzTur88, TheCarDecon06} and~\cite{HoefIl12}.

\section{Whitham approach: setup, leading order solution}

As is very common, see e.g.~the recent review~\cite{ElHoefRev16} and references therein, it is convenient to represent the 2dNLS eq.~(\ref{eq:2dNLS}) in the form of Euler type hydrodynamics equations for the ``density" and ``velocity". For this purpose, we express $\Psi = \sqrt\rho e^{i\Theta}$, with $\rho$ and $\Theta$ real. We will employ the version of Whitham theory~\cite{Whitham74} involving singular perturbations and multiple scales~\cite{Ab2011}. This means we will consider an $\epsilon$-expansion for $\epsilon \ll 1$ of the solution to 2dNLS equation of the form

\be
\Psi = \rho^{1/2}(\theta, x, y, t; \epsilon)e^{i\Theta(\theta, x, y, t; \epsilon)},   \label{eq:2.1}   
\ee

\ni where we impose the following condition for the fast phase $\theta$:   

\be
\prt_x\theta = \frac{k}{\epsilon},  \qquad  \prt_y\theta = \frac{l}{\epsilon},  \qquad   \prt_t\theta = -\frac{\omega}{\epsilon},  \label{eq:2.2}   
\ee
$k$, $l$, $\omega$ are slowly varying quantities. From the definition (\ref{eq:2.2}), it follows immediately that the slow variables satisfy conservation of waves

\be
\prt_tk + \prt_x\omega = 0,   \qquad  \prt_tl + \prt_y\omega = 0,   \qquad  \prt_xl = \prt_yk.   \label{eq:klom}   
\ee
These are important fundamental and simple Whitham equations -- in physical variables. The last equation is a compatibility condition that must be imposed initially; it then remains valid for all time.

We will find the other Whitham equations as secularity conditions enforced by the periodic functions of $\theta$, normalized to period one. Introducing

$$
q = \frac{l}{k},
$$
one can rewrite eqs.~(\ref{eq:klom}) as 

\be
\prt_tk + \prt_x\omega = 0,   \qquad  k\prt_tq + (\prt_y-q\prt_x)\omega = 0,   \qquad  \prt_xq = \frac{(\prt_y-q\prt_x)k}{k}.   \label{eq:kqom}   
\ee
We refer to the first two of these equations as kinematic equations; the last equation is a constraint as we said above.  

The hydrodynamic velocities $u$ and $v$ are introduced as 

\be
u = \epsilon\prt_x\Theta,   \qquad   v = \epsilon\prt_y\Theta.    \label{eq:3.3}    
\ee
The imaginary part of the equation resulting from the transformation (\ref{eq:2.1}) of eq.~(\ref{eq:2dNLS}) is the analog of the hydrodynamic conservation of mass,


\be
\epsilon\prt_t\rho + \epsilon(\prt_x(2\rho u) + s\prt_y(2\rho v)) = 0,  \label{eq:rho}   
\ee
while its real part can be written as


\be
\epsilon\prt_t\Theta + \rho + u^2 + sv^2 = \epsilon^2\left(\frac{\prt_{xx}\rho+s\prt_{yy}\rho}{2\rho} - \frac{(\prt_x\rho)^2+s(\prt_y\rho)^2}{4\rho^2}\right).  \label{eq:3.5}  
\ee
Unlike the imaginary part, it contains the time evolution of the total phase $\Theta$ in addition to the hydrodynamic density $\rho$ and velocities $u$, $v$. This last equation is necessary to use in order to determine the total phase $\Theta$ of the 2d NLS solution. Otherwise, it is convenient to use its $x$- and $y$-derivatives,

\be
\epsilon(\prt_tu + \prt_x(\rho+u^2+sv^2)) = \epsilon^3\prt_x\left(\frac{\prt_{xx}\rho+s\prt_{yy}\rho}{2\rho} - \frac{(\prt_x\rho)^2+s(\prt_y\rho)^2}{4\rho^2}\right).  \label{eq:u}  
\ee

\be
\epsilon(\prt_tv + \prt_y(\rho+u^2+sv^2)) = \epsilon^3\prt_y\left(\frac{\prt_{xx}\rho+s\prt_{yy}\rho}{2\rho} - \frac{(\prt_x\rho)^2+s(\prt_y\rho)^2}{4\rho^2}\right).  \label{eq:v}    
\ee
The variables $u, v$ are not independent; they satisfy the constraint 

\be
\prt_xv = \prt_yu.   \label{eq:uv}
\ee

Using multiple scales, we calculate the derivatives as the sum of fast and slow derivatives,
$$
\epsilon\prt_xf = (k\prt_{\theta} + \epsilon\tilde\prt_x)f,  \qquad  \epsilon\prt_yf = (l\prt_{\theta} + \epsilon\tilde\prt_y)f,  \qquad   \epsilon\prt_tf = (-\omega\prt_{\theta} + \epsilon\tilde\prt_t)f,
$$
where $\prt_{\theta}/\epsilon$ is the fast derivative and $\tilde\prt_x$, $\tilde\prt_y$ and $\tilde\prt_t$ are the slow space and time derivatives at fixed $\theta$, for $f=\rho$, $u$ or $v$. We also denote $f' \equiv \prt_\theta f$ here and further on. Then we expand in $\epsilon$:

\[ \rho = \rho_0 + \epsilon\rho_1 + \dots, ~~u = u_0 + \epsilon u_1 + \dots,  ~~v = v_0 + \epsilon v_1 + \dots. \]   
 
\ni However, since $\Theta$ itself is a fast variable, i.e.~is of order $1/\epsilon$, the leading order of the expansion of its derivatives will have forms
$$
\epsilon\prt_x\Theta_0 = k\Theta_0' + \alpha,   \qquad  \epsilon\prt_y\Theta_0 = kq\Theta_0' + \eta,   \qquad  \epsilon\prt_t\Theta_0 = -kV\Theta_0' - \beta, 
$$
where $\alpha$, $\eta$ and $\beta$ are additional slow variables, the leading orders of ``slow derivatives" $\epsilon\tilde\prt_x\Theta$, $\epsilon\tilde\prt_y\Theta$ and $\epsilon\tilde\prt_t\Theta$, respectively. This prescription is in fact equivalent to having two fast phases, see \cite{rNLS} for details. Note that, while $\Theta_0 \sim 1/\epsilon$, the derivative $\prt_\theta\Theta_0 \equiv \Theta_0' \sim 1$. 
\par At the leading order eq.~(\ref{eq:rho}) yields

$$
-\omega\rho_0' + 2k(\rho_0u_0)' + 2skq(\rho_0v_0)' = 0,
$$

\ni and its first integral is

\be
u_0 + sqv_0 = \frac{\omega}{2k} - \frac{C_s}{2\rho_0},  \label{eq:Im0}   
\ee

\ni where an integration `constant' $C_s$~\footnote{$C_s$ and other `constants' of integration in $\theta$ appearing later are slow variables depending on $x, y$ and $t$.} is introduced. The leading order of eq.~(\ref{eq:uv}), $v_0'=qu_0'$, yields $v_0$ in terms of $u_0$,

\be
v_0 = qu_0 + p,   \label{eq:v0}
\ee
where we introduced another integration slow variable $p$. Using this, eq.~(\ref{eq:Im0}) allows one to express $u_0$ in terms of $\rho_0$,

\be
u_0 = \frac{V}{2} - \frac{C_s}{2(1+sq^2)\rho_0}.   \label{eq:u0}
\ee
We will use the slow variables $V$ and $V_y$, 

\be
V = \frac{\omega/k - 2sqp}{1+sq^2},   \qquad qV_y = \frac{\omega}{k} - V = sq(qV + 2p).   \label{eq:VVy}
\ee

The leading order of eq.~(\ref{eq:u}) is also a total derivative in $\theta$ which integrates to
\be
\rho_0 + u_0^2 - \frac{\omega}{k}u_0 + sv_0^2 + H_u = k^2(1+sq^2)\left(\frac{\rho_0''}{2\rho_0} - \frac{(\rho_0')^2}{4\rho_0^2}\right),   \label{eq:Re0}   
\ee
where $H_u$ is another integration `constant'. Using eqs.~(\ref{eq:v0}) and (\ref{eq:u0}), this can be rewritten as 

$$
k^2(1+sq^2)(2\rho_0\rho_0'' - (\rho_0')^2) = 4\rho_0^3 - 4(1+sq^2)\Omega\rho_0^2 + \frac{C_s^2}{1+sq^2},  
$$
after introducing the slow variable $\Omega$ by $\Omega = V^2/4 - H_u$. If we let

\be
\rho_0 = (1+sq^2)\mu_0,   \label{eq:mu0}
\ee
the factor $(1+sq^2)$ can be removed from the previous equation and it becomes

\be
k^2(2\mu_0\mu_0'' - (\mu_0')^2) = 4\mu_0^3 - 4\Omega\mu_0^2 + C_0^2,   \label{eq:Re0mu}
\ee
where
\[C_0=C_s/(1+sq^2)^2.\] 
This is an ODE for an elliptic function $\mu_0$ (e.g.~assume $(\mu_0')^2=F(\mu_0)$) and one can verify that, as a consequence, $\mu_0$ satisfies

\be
2k^2(\mu_0')^2 = 4\left(\mu_0^3-2\Omega\mu_0^2+2C_1\mu_0-\frac{C_0^2}{2}\right) = 4(\mu_0-\lambda_1)(\mu_0-\lambda_2)(\mu_0-\lambda_3),  \label{eq:2.10}  
\ee
with another integration constant (slow variable) $C_1$. The general solution $\mu_0$ is 

\be
\mu_0 = a - b\;\text{cn}^2\left(2K(m)(\theta - \theta_*); m\right),   \label{eq:2.11}   
\ee
where $\theta_*$ is another integration constant (slow variable in general), we take $\lambda_1\le\lambda_2\le\lambda_3$ and then

\be
a = \lambda_2,  \qquad b = (\lambda_2-\lambda_1),  \qquad  m = \frac{\lambda_2-\lambda_1}{\lambda_3-\lambda_1}.   \label{eq:2.12}   
\ee
We have the following relations among the parameters (slow variables):

\be
e_1 \equiv \lambda_1 + \lambda_2 + \lambda_3 = 2\Omega,  \quad  e_2 \equiv \lambda_1\lambda_2+\lambda_2\lambda_3+\lambda_3\lambda_1 = 2C_1,  \quad e_3\equiv \lambda_1\lambda_2\lambda_3 = \frac{C_0^2}{2}.  \label{eq:2.13} 
\ee
The normalization of the elliptic function with fixed unit period in $\theta$ implies that

\be
k^2 = \frac{b}{8mK^2(m)} = \frac{\lambda_3-\lambda_1}{8K^2(m)},   \label{eq:2.14}   
\ee
where $K(m)$ is the first complete elliptic integral. Also it follows that the leading order hydrodynamic velocity is 
\be
u_0 = \frac{V}{2} - \frac{C_0}{2\mu_0} = \frac{V}{2} - \sigma\frac{\sqrt{2\lambda_1\lambda_2\lambda_3}}{2\mu_0},   \qquad \sigma \equiv \text{sign}(C_0),  \label{eq:3.6}   
\ee
in terms of the introduced roots of the cubic $\lambda_i$, $i=1,2,3$. The above has the same functional form as the well-known leading order solution for 1d NLS, see, however, Remark 1 in the previous section. Formula (\ref{eq:Phase}) for the total phase is derived using eq.~(\ref{eq:3.5}), together with eqs.~(\ref{eq:3.6}), (\ref{eq:v0}) and (\ref{eq:Re0mu}), in order to find the following expression for $\prt_t\Theta_0$:

\be
\epsilon\prt_t\Theta_0 = \frac{(V+qV_y)C_0}{2\mu_0(\theta; x,y,t)} - \frac{gR_2}{4} - sp(qV+p).    \label{eq:Phase_t}
\ee	
The leading order of eq.~(\ref{eq:3.5}) reads
\be
\epsilon\prt_t\Theta_0 + (1+sq^2)\mu_0 + u_0^2 + sv_0^2 = k^2(1+sq^2)\left(\frac{\mu_0''}{2\mu_0} - \frac{(\mu_0')^2}{4\mu_0^2}\right).  \label{eq:3.5'}  
\ee
Using eqs.~(\ref{eq:3.6}), (\ref{eq:v0}) and (\ref{eq:Re0mu}), this yields eq.~(\ref{eq:Phase_t}). Then the total phase $\Theta_0$ is restored by integration (up to an irrelevant constant phase) from its known partials $\epsilon\prt_x\Theta_0=u_0$, $\epsilon\prt_y\Theta_0=qu_0+p$ and $\epsilon\prt_t\Theta_0$ and eventually eq.~(\ref{eq:Phase}) is obtained.

\par At this stage we expect that the variables $\lambda_1,\lambda_2,\lambda_3,V, q$ and $p$ are going to be the key dependent variables. Knowledge of these variables allows us to reconstruct $\rho_0, u_0$ and $\theta$ and hence $\Theta_0$ and the leading order approximation for $\Psi$. These variables are similar to those that arise in the study of two dimensional KP-type systems \cite{ABW-KP,ABR}. If $C_0=0$ in eq.~(\ref{eq:3.6}), which can be the case if and only if the basic variable $\lambda_1=0$ (i.e.~the smallest of the three $\lambda$-s), then we get a solution with simple phase. 

\section{Whitham system for the 2dNLS equation}   

This section presents details of derivation of 2d NLS Whitham equations in `physical variables'; as noted above, the basic variables are $V$, $\lambda_1$, $\lambda_2$, $\lambda_3$, $q$ and $p$.

The next order ($\sim\epsilon$) of eq.~(\ref{eq:rho}) reads~\footnote{Here and further on we write $\prt_x$ for $\tilde\prt_x$, $\prt_y$ for $\tilde\prt_y$ and $\prt_t$ for $\tilde\prt_t$ but they denote the derivatives at fixed $\theta$ in all equations at order $\epsilon$ while they have usual meaning in exact equations.}

\be
-\omega\rho_1' + 2k[\rho u + sq\rho v]_1' + \prt_t\rho_0 + \prt_x(2\rho_0u_0) + \prt_y(2s\rho_0v_0) = 0,  \label{eq:rho1}   
\ee
where here and below the notation $[\dots]_1$ means that the part of expression in brackets at order $\epsilon$ is taken. Requiring the first-order corrections $\rho_1$, $u_1$ and $v_1$ to not grow but be periodic in $\theta$, and integrating eq.~(\ref{eq:rho1}) over the period, we find a secularity condition

$$
\prt_t\overline\rho_0 + \prt_x(\overline{2\rho_0u_0}) + \prt_y(\overline{2s\rho_0v_0}) = 0,
$$
or, after using the relation $v_0=qu_0+p$,

\be
\prt_t\overline\rho_0 + \prt_x(\overline{2\rho_0u_0}) + \prt_y(\overline{2s\rho_0(qu_0+p)}) = 0.   \label{eq:S1}
\ee
This is the averaged mass conservation law in the (2+1)-dimensional case.
\par Next we use the combination of eq.~(\ref{eq:u}) and eq.~(\ref{eq:rho}): $2\rho\cdot($\ref{eq:u}$) + 2u\cdot($\ref{eq:rho}$)$, to get the $x$-component of (hydrodynamic) momentum conservation law for 2dNLS, 

\be
\epsilon\prt_t(2\rho u) + \epsilon\prt_x(\rho^2+4\rho u^2) + \epsilon\prt_y(4s\rho uv) =  \epsilon^3\prt_x\left(\prt_{xx}\rho - \frac{(\prt_x\rho)^2}{\rho}\right) + \epsilon^3s\prt_y\left(\prt_{xy}\rho - \frac{\prt_x\rho\prt_y\rho}{\rho}\right).   \label{eq:Px}  
\ee
To get it we also used  the identity

$$
2\rho\prt_x(v^2) + 4u\prt_y(\rho v) = \prt_y(4\rho uv)
$$
which follows from eq.~(\ref{eq:uv}). The leading order of eq.~(\ref{eq:Px}) does not contain new information being a combination of eqs.~(\ref{eq:Im0}) and (\ref{eq:Re0}). At order $\epsilon$, eq.~(\ref{eq:Px}) reads:




$$
-2\omega[\rho u]_1' + k[\rho^2 + 4\rho u^2]_1' + skq[4\rho uv]_1' - k\left[\prt_{xx}\rho - \frac{(\prt_x\rho)^2}{\rho} + sq\left(\prt_{xy}\rho - \frac{\prt_x\rho\prt_y\rho}{\rho}\right)\right]_1' +
$$

\be
+ \prt_t(2\rho_0u_0) + \prt_x(\rho_0^2+4\rho_0u_0^2) + s\prt_y(4\rho_0u_0v_0) - \prt_x\left(k^2\left(\rho_0'' - \frac{(\rho_0')^2}{\rho_0}\right)\right) - s\prt_y\left(k^2q\left(\rho_0'' - \frac{(\rho_0')^2}{\rho_0}\right)\right) = 0.  \label{eq:Px1}   
\ee
Integrating eq.~(\ref{eq:Px1}) over the period in $\theta$, one finds another secularity condition,

$$
\prt_t(\overline{2\rho_0u_0}) + \prt_x(\overline{\rho_0^2+4\rho_0u_0^2}) + s\prt_y(\overline{4\rho_0u_0v_0}) + \prt_x\left(\overline{k^2\frac{(\rho_0')^2}{\rho_0}}\right) + s\prt_y\left(\overline{k^2q\frac{(\rho_0')^2}{\rho_0}}\right) = 0,  
$$
or, using the relation $v_0=qu_0+p$,

\be 
\prt_t(\overline{2\rho_0u_0}) + \prt_x\left(\overline{\rho_0^2+4\rho_0u_0^2 + k^2\frac{(\rho_0')^2}{\rho_0}}\right) + s\prt_y\left(q\overline{\left(4\rho_0u_0^2 + k^2\frac{(\rho_0')^2}{\rho_0}\right)} + 2p\cdot\overline{2\rho_0u_0}\right) = 0.  \label{eq:S2}
\ee
This is the $x$-component of averaged momentum conservation law for 2dNLS equation. Similarly, combining eqs.~(\ref{eq:v}) and (\ref{eq:rho}), the $y$-component of the momentum equation reads


\be
\epsilon\prt_t(2\rho v) + \epsilon\prt_y(\rho^2+4s\rho v^2) + \epsilon\prt_x(4\rho uv) =  \epsilon^3s\prt_y\left(\prt_{yy}\rho - \frac{(\prt_y\rho)^2}{\rho}\right) + \epsilon^3\prt_x\left(\prt_{xy}\rho - \frac{\prt_x\rho\prt_y\rho}{\rho}\right),   \label{eq:Py1}  
\ee
with the help of an identity following from eq.~(\ref{eq:uv}),

$$
2\rho\prt_y(u^2) + 4v\prt_x(\rho u) = \prt_x(4\rho uv).
$$
The corresponding secularity equation also follows: 

$$
\prt_t(\overline{2\rho_0v_0}) + \prt_y(\overline{\rho_0^2+4s\rho_0v_0^2}) + \prt_x(\overline{4\rho_0u_0v_0}) + s\prt_y\left(q^2\overline{k^2\frac{(\rho_0')^2}{\rho_0}}\right) + \prt_x\left(q\overline{k^2\frac{(\rho_0')^2}{\rho_0}}\right) = 0,  
$$
or, using the relation $v_0=qu_0+p$,

$$ 
\prt_t(q\cdot\overline{2\rho_0u_0} + 2p\cdot\overline{\rho_0}) + \prt_x\left(q\overline{\left(4\rho_0u_0^2 + k^2\frac{(\rho_0')^2}{\rho_0}\right)} + 2p\cdot\overline{2\rho_0u_0}\right) +
$$

\be
+ \prt_y\left(\overline{\rho_0^2+sq^2\left(4\rho_0u_0^2 + k^2\frac{(\rho_0')^2}{\rho_0}\right)} + 4sqp\cdot\overline{2\rho_0u_0} + 4sp^2\cdot\overline{\rho_0}\right) = 0.  \label{eq:S2v}
\ee
This is the $y$-component of the (averaged) momentum conservation law. 

\par We need one more secularity condition to close the system. We derive it from the energy conservation law. In hydrodynamic variables, the energy is (in our normalization for 2dNLS)

\be
H_{2d} = 2\rho^2 + 4\rho(u^2 + sv^2) + \frac{\epsilon^2((\prt_x\rho)^2 + s(\prt_y\rho)^2)}{\rho}.   \label{eq:H2d}
\ee
Differentiating this expression w.r.t.~time and using the known expressions for $\prt_t\rho$, $\prt_tu$ and $\prt_tv$ via the terms from eqs.~(\ref{eq:rho}), (\ref{eq:u}) and (\ref{eq:v}) not containing $\prt_t$, we get the energy equation in the form of a (2+1)-dimensional conservation law:




{\small $$
\epsilon\prt_t\left(2\rho^2 + 4\rho(u^2 + sv^2) + \frac{\epsilon^2((\prt_x\rho)^2 + s(\prt_y\rho)^2)}{\rho}\right) + \epsilon\prt_x[8\rho u(\rho + u^2 + sv^2)] + \epsilon\prt_y[8s\rho v(\rho + u^2 + sv^2)] =
$$}

$$
= 4\epsilon^3\prt_x\left\{u\left[\prt_{xx}\rho + s\prt_{yy}\rho - \frac{3(\prt_x\rho)^2+s(\prt_y\rho)^2}{2\rho}\right] - sv\frac{\prt_x\rho\prt_y\rho}{\rho} - \prt_x\rho[\prt_xu+s\prt_yv]\right\} +
$$

\be
+ 4\epsilon^3\prt_y\left\{sv\left[\prt_{xx}\rho + s\prt_{yy}\rho - \frac{(\prt_x\rho)^2+3s(\prt_y\rho)^2}{2\rho}\right] - su\frac{\prt_x\rho\prt_y\rho}{\rho} - s\prt_y\rho[\prt_xu+s\prt_yv]\right\}.   \label{eq:EC}
\ee
Its leading order is again a total derivative in $\theta$; integrating, we obtain another consequence of the basic leading order ODEs, so we do not show it. The next order (order $\epsilon$) of the energy equation (\ref{eq:EC}) is a long equation which we do not display; upon integration over the period, the (total) $\theta$-derivatives integrate to zero and we obtain the remaining secularity condition,

{\small $$
\prt_t\left(\overline{2\rho_0^2+4\rho_0(u_0^2+sv_0^2)+(1+sq^2)k^2\frac{(\rho_0')^2}{\rho_0}}\right) + \prt_x[\overline{8\rho_0u_0(\rho_0 + u_0^2 + sv_0^2)}] + \prt_y[\overline{8s\rho_0v_0(\rho_0 + u_0^2 + sv_0^2)}] =
$$}

{\small$$
= 4\prt_x\left\{-\overline{k^2\frac{(\rho_0')^2}{\rho_0}((3+sq^2)u_0/2 + sqv_0)} - \overline{k^2\rho_0'((2+sq^2)u_0'+sqv_0')}\right\} + 
$$

\be
+ 4\prt_y\left\{-\overline{k^2\frac{(\rho_0')^2}{\rho_0}(squ_0 + s(1+3sq^2)v_0/2)} - \overline{k^2\rho_0'(squ_0'+s(1+2sq^2)v_0')}\right\}.   \label{eq:E1av}
\ee}
Using $v_0=qu_0+p$, this becomes

{\small$$
\prt_t\left(\overline{2\rho_0^2+(1+sq^2)(4\rho_0u_0^2+\frac{k^2(\rho_0')^2}{\rho_0}) + 4sqp\cdot2\rho_0u_0 + 4sp^2\rho_0}\right) + 
$$

$$
+ \prt_x[\overline{8\rho_0^2u_0 + (1+sq^2)(8\rho_0u_0^3 + 6\frac{k^2(\rho_0')^2}{\rho_0}u_0 + 8k^2\rho_0'u_0') + 4sqp(4\rho_0u_0^2+\frac{k^2(\rho_0')^2}{\rho_0}) + 4sp^2\cdot2\rho_0u_0}] +        
$$

$$
+\prt_y[\overline{8sq\rho_0^2u_0 + sq(1+sq^2)(8\rho_0u_0^3 + 6\frac{k^2(\rho_0')^2}{\rho_0}u_0 + 8k^2\rho_0'u_0') +}
$$

\be
\overline{+ 8sp\rho_0^2 + 2s(1+3sq^2)p(4\rho_0u_0^2+\frac{k^2(\rho_0')^2}{\rho_0}) + 4s^2p^2(3q\cdot2\rho_0u_0 + 2p\rho_0)}] = 0.         \label{eq:S3}
\ee}
To rewrite our obtained system in a more compact form we denote

$$
g = 1+sq^2, \qquad Q = \overline{\mu_0},  \qquad  P = \overline{2\mu_0u_0},  \qquad  B = \overline{4\mu_0u_0^2+\frac{k^2(\mu_0')^2}{\mu_0}},  \qquad N = \overline{\mu_0^2}, 
$$

\be
H \equiv H_{2d} = g(2N + B) + 4sqpP + 4sp^2Q,  \qquad  J = \overline{8\mu_0^2u_0} + \overline{8\mu_0u_0^3 + 6\frac{k^2(\mu_0')^2}{\mu_0}u_0 + 8k^2\mu_0'u_0'}.   \label{eq:svdef}
\ee
Using the leading order solution from previous section and elliptic function identities from appendix A, we find 

\be
Q=\int_0^1\mu_0(\theta) d\theta = a + b\left(\frac{E(m)}{mK(m)} - \frac{1-m}{m}\right) = \lambda_3 - (\lambda_3-\lambda_1)\frac{E}{K},   \label{eq:Ql}  
\ee
where $K=K(m)$ and $E=E(m)$ are the first and second complete elliptic integrals, respectively;

$$
P = VQ - C_0, \qquad B = V^2Q - 2VC_0 + \frac{4(2C_1-\Omega Q)}{3},   \qquad  N = \frac{2(2\Omega Q - C_1)}{3},  
$$

$$
\overline{2\mu_0^2u_0} = VN - C_0Q,  \qquad \overline{8\mu_0u_0^3 + 6\frac{k^2(\mu_0')^2}{\mu_0}u_0 + 8k^2\mu_0'u_0'} = V^3Q - 3V^2C_0 + 4V(2C_1-\Omega Q) + 4C_0(Q-\Omega),
$$

\be
J = V(B + \frac{8}{3}(\Omega Q + C_1)) - (V^2 + 4\Omega)C_0 = V\left(V^2 + \frac{4\Omega}{3}\right)Q - (3V^2 + 4\Omega)C_0 + \frac{16VC_1}{3}.   \label{eq:svexpr}
\ee
Then the complete Whitham PDE system obtained has two kinematic equations from equations (\ref{eq:klom}),(\ref{eq:kqom}) and (\ref{eq:VVy}):

\be
\prt_tk + \prt_x(k(V+qV_y)) = 0,   \qquad V_y \equiv s(qV+2p),   \label{eq:(k)}  
\ee

\be
\prt_tq + (V+qV_y)\prt_xq + (\prt_y-q\prt_x)(V+qV_y) = 0,   \label{eq:(q)}
\ee
and four equations from secularity conditions:
\be
\prt_t(gQ) + \prt_x(gP) + \prt_y(sg(qP + 2pQ)) = 0,   \label{eq:Q}   
\ee

\be
\prt_t(gP) + \prt_x[g(B + gN)] + \prt_y[sg(qB + 2pP)] = 0,  \label{eq:P}   
\ee

\be
\prt_t[g(qP + 2pQ)] + \prt_x[g(qB + 2pP)]  + \prt_y[g(gN + sq^2B + 4sqpP + 4sp^2Q)] = 0,  \label{eq:Py}   
\ee

\be
\prt_t(gH) + \prt_x[g(gJ + 4sqpB + 4sp^2P)] + \prt_y[sg(qgJ + 8gpN + 2(1+3sq^2)pB + 12sqp^2P + 8sp^3Q)] = 0.   \label{eq:E}
\ee
Besides, there are two constraints (non-dynamical equations). One is the consistency condition, see eq.~(\ref{eq:kqom}),

\be
\prt_xq = \frac{(\prt_y-q\prt_x)k}{k}.   \label{eq:kq}
\ee
The other important constraint for $p$ is obtained from the leading order of eq.~(\ref{eq:uv}) by integrating over the period,

$$
\prt_x(q\overline u_0 + p) = \prt_y\overline u_0.
$$
Then, using eq.~(\ref{eq:3.6}), 

\be
\prt_xp = D_y\overline u_0 - \overline u_0\frac{D_yk}{k} = D_y\left(\frac{V}{2} - \overline{\frac{C_0}{2\mu_0}}\right) - \left(\frac{V}{2} - \overline{\frac{C_0}{2\mu_0}}\right)\frac{D_yk}{k}.    \label{eq:px0}
\ee
The average of $C_0/\mu_0$ can be expressed in terms of the third complete elliptic integral $\Pi(-(\lambda_2-\lambda_1)/\lambda_1, m)$, see eq.~(\ref{eq:A7}) in appendix \ref{a:El}. Then, using eqs.~(\ref{eq:mPi}) and (\ref{eq:gPi}) for the derivatives of $\Pi$-function with respect to its arguments, we derive from eq.~(\ref{eq:px0}) the expression

{\small\be
2\prt_xp = D_yV - V\frac{D_yk}{k} - \frac{QC_0}{2}\left( \frac{D_y\lambda_1}{\lambda_1\lambda_{21}\lambda_{31}} - \frac{D_y\lambda_2}{\lambda_2\lambda_{21}\lambda_{32}} + \frac{D_y\lambda_3}{\lambda_3\lambda_{32}\lambda_{31}} \right) + \frac{C_0}{2}\left( \frac{D_y\lambda_1}{\lambda_{21}\lambda_{31}} - \frac{D_y\lambda_2}{\lambda_{21}\lambda_{32}} + \frac{D_y\lambda_3}{\lambda_{32}\lambda_{31}} \right),   \label{eq:pxl}
\ee}
where we denoted $\lambda_{ij}=\lambda_i-\lambda_j$. Eq.~(\ref{eq:pxl}) complements the Whitham system and will be used in the next section.

\par All the involved dependent variables here are explicitly expressed in terms of the six basic variables $V$, $\lambda_j, j=1,2,3,$ $q,p$ using the equations preceding the listed system and recalling the relations (\ref{eq:2.13}) and (\ref{eq:2.14}); they are kept only to make the equations concise. We remark that while this form of the Whitham system has some features of the one used in~\cite{rNLS} for the radial reduction (rNLS) of 2d NLS equation, it has additional $q$ and $p$ variables and more equations, respectively. Eqs.~(\ref{eq:(k)})--(\ref{eq:E}) together with constraints (\ref{eq:kq}) and (\ref{eq:pxl}) comprise a complete system of Whitham-2dNLS equations. As indicated above, the first two are from kinematic relations and the next four of the six Whitham equations (``mass", two components of ``momentum" and ``energy") have been obtained as secularity conditions.

\section{Whitham equations in Riemann-type variables}

We derive our final 2dNLS-Whitham system (\ref{eq:Wr-eq})--(\ref{eq:px-eq}) from the system of the previous section in three major steps. First, as an intermediate step described in appendix B, we derive eqs.~(\ref{eq:tk})--(\ref{eq:qp}) from eqs.~(\ref{eq:(k)})--(\ref{eq:E}). Second, we change four of the six basic variables and combine the equations of appendix B to obtain the four equations (\ref{eq:Wr}) for the Riemann-type variables $r_j, j=1,\dots,4$. Third, we simplify the system of eqs.~(\ref{eq:Wr}), (\ref{eq:tq}), (\ref{eq:qp}) and constraints (\ref{eq:tkq}), (\ref{eq:pxl}) to get the final result. The details for the second and third steps are provided in appendix \ref{a:Rv}. 
\par We are guided by the known 1dNLS-Whitham system in Riemann invariants. The one-dimensional (1d) NLS equation (eq.~(\ref{eq:2dNLS}) without $\prt_{yy}$-term) is known~\cite{GK87, Pa87, ElKr95, HoeferEtAl06} to have Whitham equations that are diagonal in the spatial and temporal derivatives of the Riemann variables $\{r_j, j=1,\dots, 4\}$,

\be
\prt_tr_j + v_j\prt_xr_j = 0,    \qquad  v_j = V + \frac{k}{4\prt_jk},   \quad \prt_j \equiv \frac{\prt}{\prt r_j},  \label{eq:4.1}   
\ee
where the relation with variables $V$, $\lambda_j$, $j=1,2,3$, is, in our normalization,

{\small $$
V = \frac{r_1+r_2+r_3+r_4}{4},    \qquad   \lambda_1 = \frac{(r_1+r_4-r_2-r_3)^2}{32},  
$$

\be
\lambda_2 = \frac{(r_2+r_4-r_1-r_3)^2}{32},   \qquad  \lambda_3 = \frac{(r_3+r_4-r_1-r_2)^2}{32}.   \label{eq:4.3}   
\ee}

\ni Thus, the velocities $v_j$ are determined by log-derivatives of $k$ w.r.t.~the Riemann variables. The log-derivatives of $k$ are found from formula (\ref{eq:2.14}) and definitions (\ref{eq:4.3}), which imply 


\be
m = \frac{\lambda_2-\lambda_1}{\lambda_3-\lambda_1} = \frac{(r_2-r_1)(r_4-r_3)}{(r_3-r_1)(r_4-r_2)}   \label{eq:4.5}   
\ee

\ni and 

{\small $$
\frac{\prt_1k}{k} = -\frac{1}{2(r_2-r_1)}\left(1 - \frac{r_4-r_2}{r_4-r_1}\frac{E}{K}\right),   \qquad  \frac{\prt_2k}{k} = \frac{1}{2(r_2-r_1)}\left(1 - \frac{r_3-r_1}{r_3-r_2}\frac{E}{K}\right),
$$

\be
\frac{\prt_3k}{k} = -\frac{1}{2(r_4-r_3)}\left(1 - \frac{r_4-r_2}{r_3-r_2}\frac{E}{K}\right),   \qquad  \frac{\prt_4k}{k} = \frac{1}{2(r_4-r_3)}\left(1 - \frac{r_3-r_1}{r_4-r_1}\frac{E}{K}\right).   \label{eq:4.6}   
\ee}
In the above and further on we use $K \equiv K(m)$ and $E\equiv E(m)$. We also note that the above 1d NLS Whitham system for $r_j,j=1,2,3,4$ is a closed system of equations.

\par We are going to obtain the 2dNLS Whitham equations in terms of the Riemann-type variables $r_j, j=1,2,3,4$ defined by eqs.~(\ref{eq:4.3}). The relevance of the same variables here is due to the fact that the leading order solution in $\epsilon$ has the same form for 2dNLS and for 1dNLS. But unlike the 1dNLS equation, for the 2dNLS equation the analogous four equations for $r_j, j=1,2,3,4$, must be supplemented by additional equations for $q,p$.  

\par After deriving eqs.~(\ref{eq:tk})--(\ref{eq:qp}) in appendix B from eqs.~(\ref{eq:(k)})--(\ref{eq:E}), we express all the physical slow variables in terms of variables $r_j$ using eqs.~(\ref{eq:4.3}), (\ref{eq:4.5}) and (\ref{eq:4.6}). Introduce the power functions of $r_j$,
 
$$
p_n = \sum_{j=1}^4r_j^n, \qquad n = 1, 2, \dots
$$
Then 

\be
V = \frac{p_1}{4},  \qquad 2\Omega = e_1 = \frac{1}{32}\left(3p_2 - 2\sum_{j<l}r_jr_l \right) = \frac{1}{32}(4p_2 - p_1^2),  \qquad R_2 = V^2+4\Omega = \frac{p_2}{4},    \label{eq:Ve1r}
\ee

\be
C_0 = \sigma\sqrt{2e_3} = -\frac{1}{128}\left(2p_3 - p_1p_2 + 2\sum_{j<l<m}r_jr_lr_m\right) = -\frac{1}{16}\left(\frac{p_3}{3} - \frac{p_1p_2}{4} + \frac{p_1^3}{24}\right),  \label{eq:C0r}
\ee

\be
2C_1 = e_2 = \frac{10p_4 - 8p_1p_3 - p_2^2 + 2p_1^2p_2 - 24r_1r_2r_3r_4}{(32)^2} = \frac{16p_4 - 16p_1p_3 - 4p_2^2 + 8p_1^2p_2 - p_1^4}{(32)^2},    \label{eq:e2r}   
\ee

\be
k^2 = \frac{(r_3-r_1)(r_4-r_2)}{64K^2(m)},  \qquad  m = \frac{(r_1-r_2)(r_3-r_4)}{(r_1-r_3)(r_2-r_4)},   \label{eq:k2r}
\ee

\be
Q = \frac{(r_3+r_4-r_1-r_2)^2}{32} - \frac{(r_3-r_1)(r_4-r_2)}{8}\frac{E(m)}{K(m)},   \label{eq:QrE}
\ee
and also


\be
\prt_jQ - Q\frac{\prt_jk}{k} = \frac{1}{4}r_j^2\frac{\prt_jk}{k} - \frac{p_1}{8}r_j\frac{\prt_jk}{k} + \frac{(p_1^2 - 2p_2)}{32}\frac{\prt_jk}{k}, \quad j = 1, \dots, 4.   \label{eq:dQkj}
\ee
Using the last formulas, and taking appropriate linear combinations of eqs.~(\ref{eq:tk})--(\ref{eq:ttP2}), we bring them to the form

\be  
\sum_{j=1}^4r_j^{n-1}\left(4\frac{\prt_jk}{k}Dr_j + D_xr_j\right) + W_n = 0,  \quad n = 1, \dots, 4,   \label{eq:DWr}
\ee
where all $W_n$ are such that $W_n=0$ for the 1d NLS reduction along $x$-axis (i.e.~taking $\prt_y = 0$ and $q=p=0$ in the equations),

$$
gW_1 = 4sq[Dq + VD_xq + 2D_xp] = g\cdot4sq[V\prt_xq + 2\prt_xp - D_yV],    
$$

$$
gW_2 = 4sq[2VDq + R_2D_xq + 2(Dp + VD_xp)],
$$

$$
gW_3 = (R_2-2V^2)gW_1 + 2VgW_2 + 16s[Q(qDq + VD_yq + 2D_yp) - C_0(3qD_xq + D_yq)],
$$

$$
gW_4 = (-12C_0 - VR_2 + 2V^3)gW_1 + (R_2 - 4V^2)gW_2 +  3VgW_3 + 
$$

\be
+ 16sq[PDq + 2QDp + 8C_1D_xq] + 16s[(2C_1-N-VC_0)D_yq - 2C_0D_yp],     \label{eq:W1234}
\ee

\be
D = \prt_t + V\prt_x + V_y\prt_y,   \qquad  D_x = \prt_x + sq\prt_y,   \qquad  D_y = \prt_y - q\prt_x.     \label{eq:Ds}
\ee
The four eqs.~(\ref{eq:DWr}) are equivalent to the four eqs.~(\ref{eq:Wr}) below; here the velocities $v_j$ are the 1d NLS velocities given by eqs.~(\ref{eq:4.1}) and (\ref{eq:4.6}) i.e.~eq.~(\ref{eq:vs}):

\be
\prt_tr_j + v_j\prt_xr_j + s(qv_j + 2p)\prt_yr_j + h_j(r_1,r_2,r_3,r_4,q,p) = 0,   \quad j=1,2,3,4  \label{eq:Wr}    
\ee

\be
h_j = \frac{(-1)^{j+1}\Delta_{ilm}}{4|\Delta|\prt_jk/k}[r_ir_lr_mW_1 - (r_ir_l+r_lr_m+r_mr_i)W_2 + (r_i+r_l+r_m)W_3 - W_4],  \label{eq:hj}
\ee
where

$$
|\Delta| = \prod_{j>l}^4(r_j-r_l),  \qquad  \Delta_{ilm} = (r_i-r_l)(r_l-r_m)(r_m-r_i),   \quad j\neq i\neq l\neq m\neq j  
$$
The details are given in appendix \ref{a:Rv}. Now the dynamical equations we work with are (\ref{eq:Wr}), (\ref{eq:tq}) and (\ref{eq:qp}). Note that all the quantities in eqs.~(\ref{eq:tq}), (\ref{eq:qp}) and (\ref{eq:W1234}) have explicit expressions in terms of the four $r_j$-variables, $q$ and $p$, provided by the formulas (\ref{eq:Ve1r})--(\ref{eq:QrE}), (\ref{eq:Ds}) and we recall the definitions 

\[ P=VQ-C_0, \qquad 3N=4\Omega Q - 2C_1=(R_2-V^2)Q - 2C_1,  \qquad g=1+sq^2.\]        
 
We transform eqs.~(\ref{eq:qp}) and (\ref{eq:pxl}) to a simpler form using eqs.~(\ref{eq:tq}), (\ref{eq:tkq}), (\ref{eq:pxl}) and the identities (\ref{eq:Q-j}) and (\ref{eq:Id2*}); the details are in appendix \ref{a:Rv}. The outcome is that constraint (\ref{eq:pxl}) is equivalent to eq.~(\ref{eq:px-eq}) in terms of $r$-variables and eq.~(\ref{eq:qp}) simplifies to    

\be
2Dp + VDq + 4sq\Omega D_yq + 2gD_y\Omega = 0.   \label{eq:qp-simple}
\ee
The last equation combined with eq.~(\ref{eq:tq}) to remove $Dq$-term yields eq.~(\ref{eq:p-eq}).

After some manipulations described in appendix \ref{a:Rv} eqs.~(\ref{eq:hj}) take the form

{\small$$
gh_j = sq(Dq + VD_xq + 2D_xp)(v_j - V) + s(qDq + VD_yq + 2D_yp)(r_j - v_j + V) + s\left[2qDp - 2VD_yp - \frac{(R_2+2V^2)D_yq}{3} \right] + 
$$

\be
+ 4s\left(qD_xq + \frac{D_yq}{3}\right)(v_j - V)\cdot\frac{(R_2-V^2)(r_j^2 - 2Vr_j + 2V^2-R_2) - 12C_0(r_j-V) + 32C_1}{(r_j-r_i)(r_j-r_l)(r_j-r_m)},   \label{eq:hj1}
\ee}
where one should recall eqs.~(\ref{eq:Ve1r})--(\ref{eq:e2r}). Then we use eqs.~(\ref{eq:tq}), (\ref{eq:qp-simple}) and (\ref{eq:r4id}) to get the final equation (\ref{eq:hj-eq}) from (\ref{eq:hj1}).

\par There are six equations and two constraints that must be imposed initially for six variables: $r_j, j=1,.2,3,4$ and $q,p$. Knowledge of the slow $r_j$, $q$ and $p$-variables allows us to reconstruct the leading order modulated periodic solution of the 2dNLS equation, which in suitable circumstances can describe a DSW propagating in a particular direction. The difference of our $r_j$-equations from the 1d NLS $r_j$-system is due to the additional terms $\sim\prt_yr_j$ and $h_j$. Recall that the modulated periodic solution of the 2dNLS equation has the same form as that of the 1dNLS equation. The velocities $v_j$ here are also those of 1dNLS Whitham system~\cite{GK87, Pa87, ElKr95, HoeferEtAl06}. 

\section{Linear stability analysis of traveling wave solutions}

For the system of eqs.~(\ref{eq:Wr-eq}), (\ref{eq:q-eq}) and (\ref{eq:p-eq}), consider the traveling wave solution along $x$-axis and small periodic perturbations to it i.e.~look for a linearized solution of the form

\be
r_j = \bar r_j + \rho_je^{i(\kappa x + ly - \omega t)},  \qquad q = \eta e^{i(\kappa x + ly - \omega t)},  \qquad p = \nu e^{i(\kappa x + ly - \omega t)},   \label{eq:lan}
\ee
where $\bar r_j$, $\kappa$, $l$, $\omega$ are constants and $\rho_j$, $\eta$, $\nu$ are small (constant) perturbation amplitudes. We have in linear approximation

$$
\prt_t + V\prt_x \to i(\bar V\kappa - \omega),  \qquad \prt_x \to i\kappa,  \qquad D_y \approx \prt_y \to il,   \qquad g\approx1. 
$$
Thus, the six linearized Whitham equations are

\be
i(\bar v_j\kappa - \omega)\rho_j + e^{-i(\kappa x + ly - \omega t)}h_j = 0,  \quad j=1,2,3,4  \label{eq:lrhj}
\ee

\be
(\bar V\kappa - \omega)\eta + \frac{l}{4}\sum_j\rho_j = 0,   \label{eq:lq}
\ee


\be
2(\bar V\kappa - \omega)\nu + \frac{l}{4}\sum_j(\bar r_j-2\bar V)\rho_j = 0.   \label{eq:lp1}
\ee
We linearize the quantities called $h_j$ in (\ref{eq:Wr-eq}) given by eq.~(\ref{eq:hj-eq}) and eqs.~(\ref{eq:lrhj}) take the form 





\be
(\bar v_j\kappa - \omega)\rho_j + sl\Phi_j\eta + 2sl(\bar r_j-\bar v_j)\nu = 0,  \quad j=1,2,3,4  \label{eq:lrj}
\ee
where

\be
\Phi_j = \frac{(\bar r_j-\bar V)(\bar v_j-\bar V) - \bar R_2 + \bar V^2}{3} + \bar V(\bar r_j-\bar v_j).   \label{eq:Phij}
\ee




Thus, we have a linear homogeneous algebraic $6\times6$ system for the perturbation amplitudes $\rho_j, j=1,2,3,4$, $\eta$ and $\nu$. It is solvable when its determinant is zero i.e.


\be
\left| \begin{array}{cccccc} \bar v_1\kappa - \omega & 0 & 0 & 0 & sl\Phi_1 & sl\Psi_1 \\ 0 & \bar v_2\kappa - \omega & 0 & 0 & sl\Phi_2 & sl\Psi_2 \\ 0 & 0 & \bar v_3\kappa - \omega & 0 & sl\Phi_3 & sl\Psi_3 \\ 0 & 0 & 0 & \bar v_4\kappa - \omega & sl\Phi_4 & sl\Psi_4 \\ l/4 & l/4 & l/4 & l/4 & \bar V\kappa - \omega & 0 \\ l\Pi_1 & l\Pi_2 & l\Pi_3 & l\Pi_4 & 0 & 2(\bar V\kappa - \omega) \end{array} \right| = 0,   \label{eq:ldet}
\ee
where we denoted
\be
\Psi_j = 2(\bar r_j-\bar v_j),  \qquad \Pi_j = \frac{\bar r_j-2\bar V}{4}.   \label{eq:PsiPi}
\ee
If one considers the case of longitudinal perturbations i.e.~takes $l=0$ in eq.~(\ref{eq:ldet}), then the expression simplifies,


\be
\prod_j(\bar v_j\kappa - \omega)\cdot(\bar V\kappa - \omega)^2 = 0.    \label{eq:stl}
\ee
This demonstrates the longitudinal linear stability of the traveling waves since all values of $\omega$ are real. A particularly interesting special case is that of transverse stability when one takes $\kappa=0$ in eq.~(\ref{eq:ldet}). After expanding the $6\times6$ determinant in its sixth column, then expanding each minor in its last (fifth) column and then computing the obtained (simple) $4\times4$ minors and combining similar terms, one finds the transverse dispersion relation in the form


\be
\omega^2\left\{2\omega^4 - sl^2\left[\sum_j\Psi_j\Pi_j + \frac{1}{2}\sum_j\Phi_j\right]\omega^2 + \frac{s^2l^4}{4}\left[\sum_j\Phi_j\cdot\sum_j\Psi_j\Pi_j - \sum_j\Psi_j\cdot\sum_j\Phi_j\Pi_j\right] \right\} = 0.   \label{eq:stt}
\ee
Thus, besides the two trivial modes $\omega=0$, we have two nontrivial branches -- the solutions of the quadratic equation for $\omega^2/sl^2$,



\be
\frac{\omega^2}{sl^2} = \frac{B \pm \sqrt{B^2 - 8C}}{4},    \label{eq:tsol}
\ee

\be
B = \sum_j\Psi_j\Pi_j + \frac{1}{2}\sum_j\Phi_j,   \qquad  C = \frac{1}{4}\left[\sum_j\Phi_j\cdot\sum_j\Psi_j\Pi_j - \sum_j\Psi_j\cdot\sum_j\Phi_j\Pi_j\right].   \label{eq:BC}
\ee
Note that neither $B$ nor $C$ depends on $s$ therefore $\omega^2/l^2$ is just proportional to $s$.

\par For numerical investigation of linear stability of one-dimensional traveling waves, we use the values (taken from the data of paper~\cite{rNLS})

\be
\bar r_1 = -2\sqrt2,  \qquad  \bar r_2 = 2\sqrt2,  \qquad \bar r_4 = 2\sqrt2(2\sqrt{b} - 1).     \label{eq:rfix}
\ee
The parameter $\sqrt{b} - 1$ is proportional to the dark soliton amplitude, see below in eq.~(\ref{eq:rhos-eq}). In~~\cite{rNLS}, the parameter $b-1$ had the meaning of the initial jump discontinuity of the density function $\rho$ leading to the DSW formation; here we just use the same form to parametrize the (non-modulated) traveling wave solution. We vary the two independent relevant parameters, $b$: $1 < b < \infty$ and elliptic modulus $m$: $0<m<1$. Varying more parameters should not give anything new since it corresponds to effective translation and scale transformations, symmetries of 2d NLS equation which account for the two of the four $r$-parameters involved, see e.g.~\cite{HoefIl12} for details. For a given value of elliptic modulus $m$, we then get

\be
\bar r_3 = \bar r_3(m) = \frac{\bar r_1\bar r_{42}m + \bar r_4\bar r_{21}}{\bar r_{42}m + \bar r_{21}},   \qquad  r_{jl} \equiv r_j - r_l.   \label{eq:r3m}
\ee
We use eq.~(\ref{eq:vs}) and the equations after eq.~(\ref{eq:hj-eq}) to compute $B$ and $C$ in eq.~(\ref{eq:BC}) from eqs.~(\ref{eq:Phij}) and (\ref{eq:PsiPi}).




\begin{figure}
\begin{centering}
\includegraphics[scale=0.35]{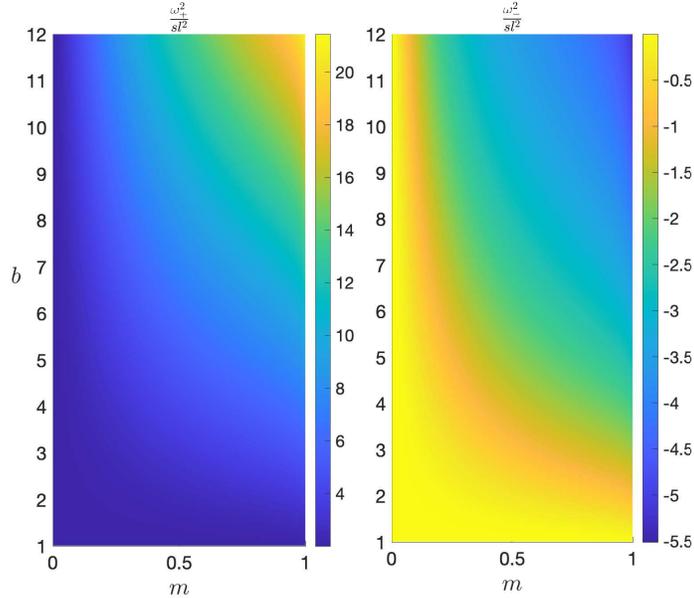}
\caption{Dispersion ratios given in eq.~(\ref{eq:tsol}). Plotted is (left) $\frac{\omega_+^2}{s l^2}$ and (right) $\frac{\omega_-^2}{s l^2}$ as a function of $b$ and $m$. Note that $\frac{\omega^2}{l^2} < 0$ indicates instability.} 
\label{whitham_disp_surfaces}
\end{centering}
\end{figure}

Unstable long wavelength modes for the family of nontrivial phase solutions (\ref{eq:rhotw})-(\ref{eq:tPhase0}) with $C_0 \neq 0$ are shown in Fig.~\ref{whitham_disp_surfaces}. The two branches of ratio $\omega^2/l^2$ in eq.~(\ref{eq:tsol}) are plotted for the ranges $m \in (0,1)$ and $b \in (1,12)$. For each sign of $s$, two modes are always real (have positive square), while the other two are imaginary (have negative square). This indicates linear transverse instability for periodic traveling waves of both elliptic ($s = 1$) and hyperbolic ($s = -1$) 2d NLS equations. This agrees with the results existing in literature, e.g.~\cite{CarterThesis, CarSeg03, TheCarDecon06}.   

\par In terms of the instability growth rates, the trend is apparent: growth rates increase with $m$ and $b$. In general, for fixed values of $(m,b)$, unstable long wavelength modes of the hyperbolic equation tend to be stronger (larger growth rate) than their elliptic counterparts (see Fig.~\ref{whitham_disp_surfaces}). Physically speaking, the most unstable nontrivial phase solutions correspond to those of dark solitons ($m \rightarrow 1$) on large backgrounds $(b \gg 1)$.  

Our direct numerics, discussed in the next subsection, for perturbed traveling wave solutions of 2d NLS equations confirm these results for small values of $l$, for which the developed Whitham theory is applicable. Furthermore, the numerics are able to describe the stability in regimes where Whitham theory does not hold, e.g.~higher frequency modes. The evolution of transversely unstable modes is also examined.

\subsection{Numerical comparisons and calculations}

The stability of 2d NLS equation eq.~(\ref{eq:2dNLS}) is now computed numerically for solution $\Psi = \rho^{1/2}(x, y, t; \epsilon)e^{i\Theta(x, y, t; \epsilon)}$, where $\rho$ is the amplitude squared and $\Theta$ is the total phase of complex function $\Psi$. We study a perturbed traveling wave solution propagating along $x$-axis. The traveling wave (leading order) solution $\Psi_0$ is defined as follows:

\be
\Psi_0 = \rho_0^{1/2}(x, t; \epsilon)e^{i\Theta_0(x, t; \epsilon)},   \label{eq:2.1''}  
\ee
where $\rho_0$ is given by eq.~(\ref{eq:2.11}) with the fast phase function $\theta(x,t;\epsilon)$ given by

\be
\theta(x,t;\epsilon) = \frac{k(x - Vt)}{\epsilon},  \qquad  k = \frac{\sqrt{r_{31}r_{42}}}{8K(m)}.   \label{eq:fphase}
\ee
In terms of $r$-variables, the general traveling wave solution along $x$-axis is now given by

\be
\rho_0 = \rho_0(x-Vt; \epsilon) = \frac{(r_{43}-r_{21})^2}{32} - \frac{r_{21}r_{43}}{8}\;\text{cn}^2\left(\frac{\sqrt{r_{31}r_{42}}(x-Vt)}{4\epsilon}; m\right),  \qquad  m = \frac{r_{21}r_{43}}{r_{31}r_{42}},   \label{eq:rhotw}   
\ee
\be
\epsilon\Theta_0 = \frac{Vx}{2} - \frac{R_2t}{4} - \frac{C_0}{2}\int_0^{x-Vt}\frac{ds}{\rho_0(s;\epsilon)},      \label{eq:tPhase0}    
\ee
with 
{\small$$
V = \frac{\sum_{j=1}^4r_j}{4},    \quad   R_2 = \frac{\sum_{j=1}^4r_j^2}{4},   \quad  C_0 = -\frac{(r_1+r_4-r_2-r_3)(r_2+r_4-r_1-r_3)(r_3+r_4-r_1-r_2)}{128}.     
$$} 
We take the $r$-parameters of eq.~(\ref{eq:rfix}) and $r_3$ determined by eq.~(\ref{eq:r3m}) in terms of elliptic modulus $m$. The elliptic modulus $m$, $0<m<1$, and the other parameter $b>1$ are varied. As mentioned earlier, the other two of the four basic parameters can be fixed by translation and scaling transformations.

\par A special case is that of traveling wave solutions with a linear phase. This occurs when $\lambda_1=0$. Then $C_0=0$ and the general expressions are (not using eq.~(\ref{eq:rfix}))

\be
\rho_0 = \frac{r_{21}^2}{8}\;\text{sn}^2\left(\frac{r_{42}(x-Vt)}{4\epsilon}; m\right),    \qquad  m = \frac{r_{21}^2}{r_{42}^2},   \label{eq:rho0c}
\ee
and
\be
\epsilon\Theta_0 = \frac{Vx}{2} - \left(\frac{V^2}{4} + \frac{r_{21}^2+r_{42}^2}{16}\right)t = \frac{(r_1+r_4)x}{4} - \left(\frac{(r_1+r_4)^2}{16} + \frac{r_{21}^2+r_{42}^2}{16}\right)t.   \label{eq:tPhase0c}
\ee
With our choice of parameters in eq.~(\ref{eq:rfix}), such solutions exist only for $b\ge 4$; then

$$
m = \frac{1}{(\sqrt{b}-1)^2},  \qquad V = 2\sqrt2(\sqrt{b}-1), \qquad r_{31} = r_{42} = 4\sqrt2(\sqrt{b}-1),    
$$
therefore

\be
\rho_0 = 4\;\text{sn}^2\left(\frac{\sqrt2(\sqrt{b}-1)(x-Vt)}{\epsilon}; m\right)   \label{eq:rho00}
\ee
and

\be
\epsilon\Theta_0 = \frac{V(x-Vt)}{2} - 2t.   \label{eq:tPhase00}
\ee

\par Another special case is the solitonic limit $m\to1$. This corresponds to $r_3\to r_2$ or $\lambda_2\to\lambda_3$. As follows from the limits of eqs.~(\ref{eq:rhotw}) and (\ref{eq:tPhase0}), this limit is described by
\be
\rho_s = \frac{(r_{42}-r_{21})^2}{32} + \frac{r_{21}r_{42}}{8}\tanh^2\theta_s  \quad\stackrel{(\ref{eq:rfix})}{=}  (\sqrt{b}-2)^2 + 4(\sqrt{b}-1)\tanh^2\theta_s,        \label{eq:rhos-eq}
\ee
and
\be
\Theta_s = \arctan\left(\frac{(r_{21}-r_{42})}{2(r_{21}r_{42})^{1/2}\tanh\theta_s}\right) + \left(\frac{r_{42}-r_{21}}{8} + \frac{V_s}{2}\right)\frac{x}{\epsilon} - \left(\frac{V_s(r_{42}-r_{21})}{8} + \frac{R_{2s}}{4}\right)\frac{t}{\epsilon},   \label{eq:sPhase-eq}
\ee
or equivalently by $\Psi_s=\rho_s^{1/2}e^{i\Theta_s}$,
\be
\Psi_s = \frac{e^{-i\phi_s(x,t)/\epsilon}}{4\sqrt2}\left[ 2(r_{21}r_{42})^{1/2}\tanh\theta_s + i(r_{21}-r_{42}) \right],   \qquad \theta_s = \frac{(r_{21}r_{42})^{1/2}}{4}\cdot\frac{(x-V_st)}{\epsilon},   \label{eq:psisf}
\ee

{\small$$
V_s = \frac{r_1+2r_2+r_4}{4} \stackrel{(\ref{eq:rfix})}{=} \sqrt{2b},  \qquad \phi_s(x,t) = -\left(\frac{r_{42}-r_{21}}{8} + \frac{V_s}{2}\right)x + \left(\frac{V_s(r_{42}-r_{21})}{8} + \frac{r_1+2r_2^2+r_4}{16}\right)t.
$$}

\par A numerical approximation of the transverse stability of solution (\ref{eq:2.1''}) was carried out. This approach is useful for verifying the Whitham theory approximations above, as well as probing regions the Whitham approach does not cover, e.g. short wavelength perturbations. We start by moving to a traveling frame of reference $z = x - V t$ in eq.~(\ref{eq:2dNLS}), which gives
\begin{equation}
\label{NLS_travel_frame}
- i \epsilon V \partial_z \Psi + i \epsilon \partial_t \Psi + \epsilon^2 \left( \partial_{zz} + s \partial_{yy} \right) \Psi - | \Psi|^2 \Psi = 0.
\end{equation}
Consider transversely-dependent perturbations of the form
\begin{equation}
\label{Numeric_perturb}
\Psi(z,y,t) = e^{i \Theta_0(z,t)} \left[ \rho_0^{1/2}(z) + \phi(z,y,t) \right] ,
\end{equation}
where $|\phi| \ll 1$ and $\Theta_0$ and $\rho_0$ are given in eqs.~(\ref{eq:tPhase0}) and (\ref{eq:rhotw}), respectively. Linearizing about the solution, we obtain the linear stability equation
\begin{align}
\label{stability_eqn}
& \epsilon V (\partial_z \Theta_0) \phi - i \epsilon V \partial_z \phi - \epsilon \partial_t \Theta_0 + i \epsilon \partial_t \phi \\ &+ \epsilon^2 \left[ \partial_{zz} \phi + 2 i (\partial_z \Theta_0)( \partial_z \phi) + i (\partial_{zz} \Theta_0) \phi - (\partial_z \Theta_0 )^2 \phi + s \partial_{yy} \phi  \right] - \rho_0 \left( \phi^*  + 2  \phi  \right)  = 0 . \nonumber
\end{align}
\begin{figure}
\begin{centering}
\includegraphics[scale=0.4]{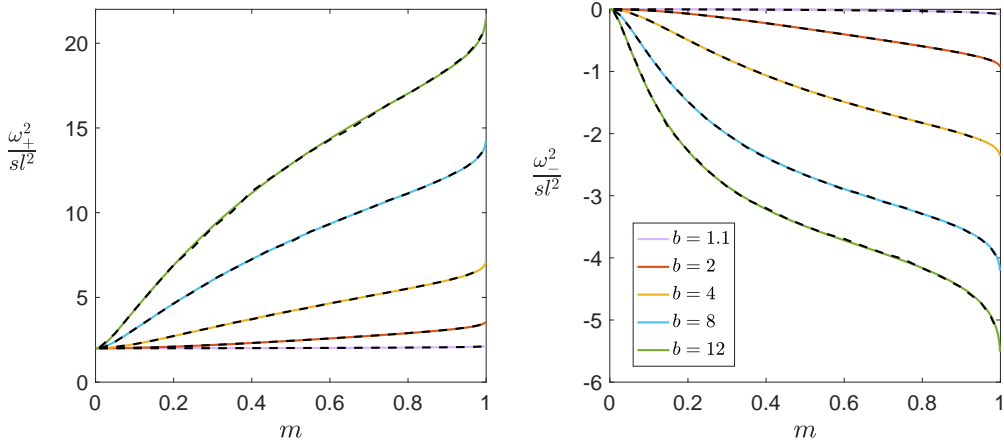}
\caption{Comparison of dispersion ratios in eq.~(\ref{eq:tsol}) versus numerical approximation. Colored curves correspond to evaluation of the function in eq.~(\ref{eq:tsol}), while  black dashed curves correspond to numerical approximations at the same values.}
\label{compare_whitham_numerics}
\end{centering}
\end{figure}
In general, this equation is periodic in $z$, with period $\epsilon/k$. The exception is the simple phase solution (\ref{eq:rho00}), which reduces to an $sn$ function and has period $2 \epsilon / k$. For equation (\ref{stability_eqn}), which is constant coefficient in $y$ and $t$, we look for exponential solutions of the form
\begin{equation}
\phi(z,y,t) = u(z) e^{i (\ell y - \omega t)} + v^*(z) e^{-i (\ell y - \omega t)} ,
\end{equation}
where $u(z + \epsilon/k) = u(z)$ and $v(z + \epsilon/k) = v(z)$. Similar to above, when $\Im \omega \not= 0$, the solution $\Psi$ is linearly unstable to transverse perturbations, with exponentially growing amplitude. The resulting eigenvalue system is 
\begin{equation}
\begin{pmatrix}
\mathcal{L}_{11} &  \mathcal{L}_{12} \\  
- \mathcal{L}_{12} &  - \mathcal{L}_{11}^* 
\end{pmatrix}
\begin{pmatrix}
u \\  v
\end{pmatrix}
 = \epsilon \omega  \begin{pmatrix}
u \\  v
\end{pmatrix}
\end{equation}
where 
\begin{align}
\mathcal{L}_{11} &= - \epsilon V (\partial_z \Theta_0) + i \epsilon V \partial_z + \epsilon \partial_t \Theta_0  - \epsilon^2 \left[ \partial_{zz} + 2 i (\partial_z \Theta_0 ) \partial_z + i \partial_{zz} \Theta_0 - (\partial_z \Theta_0)^2 - s \ell^2 \right] + 2 \rho_0 \\
\mathcal{L}_{12} &= \rho_0 
\end{align} 
Since we have periodic coefficients and eigenfunctions, all functions are expanded in their respective Fourier series. The resulting algebraic eigenvalue problem is solved by a standard eigenvalue solver, see e.g.~\cite{Yang10}. Enough Fourier modes are taken to ensure that any truncated modes are negligibly small.

\par The dispersion ratio in eq.~(\ref{eq:tsol}) was compared with a numerical approximation of the unstable modes in the long wavelength limit. The comparison is shown in Fig.~\ref{compare_whitham_numerics}. Overall, there is good agreement between the two approaches. From a numerics point of view, the most difficult region to probe is near the curve $m = \frac{1}{( \sqrt{b} - 1)^2}$, where the nontrivial solution simplifies to the simple phase solution in eqs.~(\ref{eq:rho00})-(\ref{eq:tPhase00}). Near, but not at the vacuum point, $\rho_0 \approx 0$ at $z = 0$, then the phase term in eq.~(\ref{eq:tPhase0}) becomes difficult to resolve. In contrast, the Whitham formula in eq.~(\ref{eq:tsol}) is relatively easy to evaluate for all values of $m$ and $b$ considered here.

\begin{figure}
\begin{centering}
\includegraphics[scale=0.35]{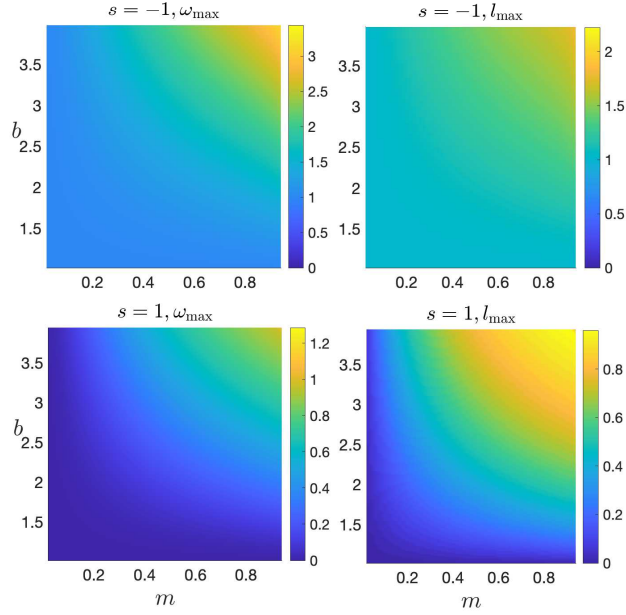}
\caption{Largest unstable mode, $\omega_{\max}$, and its corresponding transverse wavenumber, $l_{\max}$, for $\epsilon = 1$.}
\label{max_untable_mode}
\end{centering}
\end{figure}

\par Another check of the Whitham and numerics results is that of the long wavelength instability for dark solitons given in \cite{KuzTur88}. In terms of our equation (\ref{eq:2dNLS}) and solution (\ref{eq:psisf}), the elliptic ($s = 1$) transverse instability grows like $\omega^2 \sim \frac{8 l^2}{3} $ for $l \ll 1$. The Whitham and numerical approaches were found to reproduce this result.  

\begin{figure}
\begin{centering}
\includegraphics[scale=0.3]{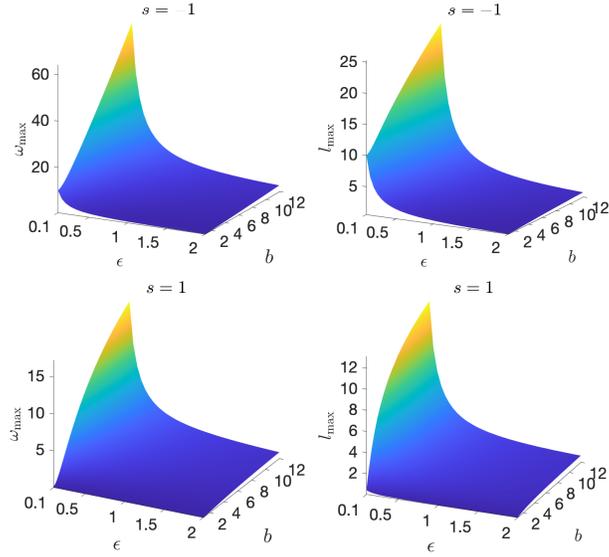}
\caption{Largest unstable mode, $\omega_{\max}$, and its corresponding transverse wavenumber, $l_{\max}$, for $m = 1/2$. The range of transverse wavenumbers scanned is $l \in [0,30]$.}
\label{max_imag_scan_eps_b_m05}
\end{centering}
\end{figure}

\par Next, the value and location of the most unstable mode is explored. Physically speaking, the most unstable mode is important since it is likely to be the dominant mode in the event of a random or localized transverse perturbation (see next subsection for simulations). To do this, first we compute $\omega_{\max} = \max_{l \in [0,10] } \omega(l)$ at wavenumber $l = l_{\max}$. The results for $\epsilon = 1$ are shown in Fig.~\ref{max_untable_mode} within the parameter regime $(b, m) \in (1,4) \times (0,1)$. The panels indicate that the strength of the instability grows as $b$ increases and $m \rightarrow 1$, similar to the long wavelength results in Fig.~\ref{whitham_disp_surfaces}. The location of the maximum instability is seen to shift to higher frequencies as $b$ increases and $m$ approaches $1$. We note that this trend appears to continue for larger values of $b$ as well. 

\par The effect of changing $\epsilon$ is shown in Fig.~\ref{max_imag_scan_eps_b_m05}, for fixed $m$. As $\epsilon$ decreases, we observe that the maximum growth rate, $\omega_{\max}$, increases and occurs at higher transverse wavenumbers, $l_{\max}$. A similar type of behavior was observed at other values of $m$.

\par Finally, we point out that in each of the comparisons between the hyperbolic ($s = -1$) and elliptic ($s = 1$) equations, the hyperbolic version has larger maximum growth rates than the elliptic case. In the long wavelength limit (see Fig.~\ref{whitham_disp_surfaces}), the hyperbolic growth rates are typically about 3--4 times larger than the elliptic ones. In the largest growth comparisons of Figs.~\ref{max_untable_mode} and \ref{max_imag_scan_eps_b_m05}, the hyperbolic growth rate is also at least three times as large as the elliptic. The wavenumber corresponding to the maximum unstable growth rate also tends to occur at higher frequencies for the hyperbolic version.

\subsubsection{Instability evolution}

In this final subsection, the evolution of perturbed traveling waves is examined. In the traveling frame of reference, the modulus and phase of periodic solution (\ref{eq:2.1''}) have different periods in $z$. To find an equation with commensurate period for simulation, we look for solutions of the form
\begin{equation}
\Psi(z,y,t) = \Phi(z,y,t) e^{i \frac{g(z)}{\epsilon}} , ~~~~~~ \partial_z g = \frac{V}{2} - \frac{C_0}{2 \rho_0(z)} 
\end{equation}
in eq.~(\ref{eq:2dNLS}), which yields
\begin{equation}
\label{NLS_phase_brought_down}
i\epsilon \prt_t\Phi  - i \epsilon V \prt_z \Phi + \left( (\prt_z g) \left[ V + 2i \epsilon \prt_z \right] - (\prt_z  g)^2 + i \epsilon (\prt_{zz} g)   \right) \Phi   + \epsilon^2(\prt_{zz}+s\prt_{yy})\Phi - |\Phi|^2\Phi = 0 ,
\end{equation}
and periodic boundary conditions in $z$: $\Phi(z + \epsilon/k,y,t) = \Phi(z,y,t)$. Numerically, all spatial derivatives are approximated by spectral Fourier methods. The equation is time-stepped using a fourth-order Runge-Kutta scheme using
the initial condition 
\begin{equation}
\label{NLS_ICs}
\Phi(z,y,0) = \rho_0^{1/2}(z) + \frac{1}{10} e^{- y^2/8} .
\end{equation}
This function represents a small localized disturbance of the traveling wave and is expected to excite transversely unstable modes, if there are any.

\begin{figure}
\begin{centering}
\includegraphics[scale=0.42]{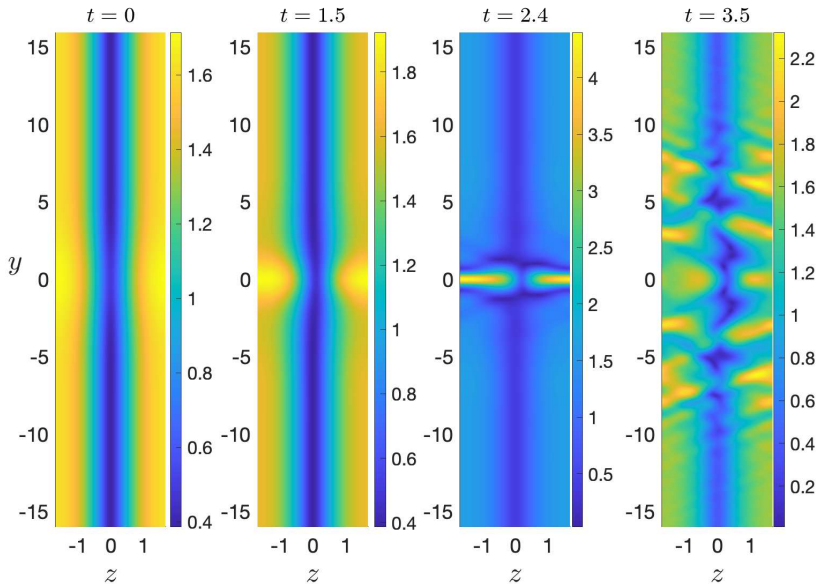}
\caption{Evolution of a perturbed traveling wave in hyperbolic ($s = -1$) NLS eq.~(\ref{NLS_phase_brought_down}). Plotted is $|\Phi(z,y,t)|$ for initial condition (\ref{NLS_ICs}) with parameters: $m = 0.75, b = 3, \epsilon = 1$ and period $1/k \approx 3.37.$}
\label{hyper_evolve}
\end{centering}
\end{figure}

\par Depending on the sign of $s$, i.e.~hyperbolic vs.~elliptic NLS, two typical instability developments are observed. In Fig.~\ref{hyper_evolve} a common instability evolution is shown for the hyperbolic NLS equation. A ``neck" type instability is observed to focus near the location of the perturbation. Subsequently, higher frequency modes radiate away from the central peak.

\begin{figure}
\begin{centering}
\includegraphics[scale=0.41]{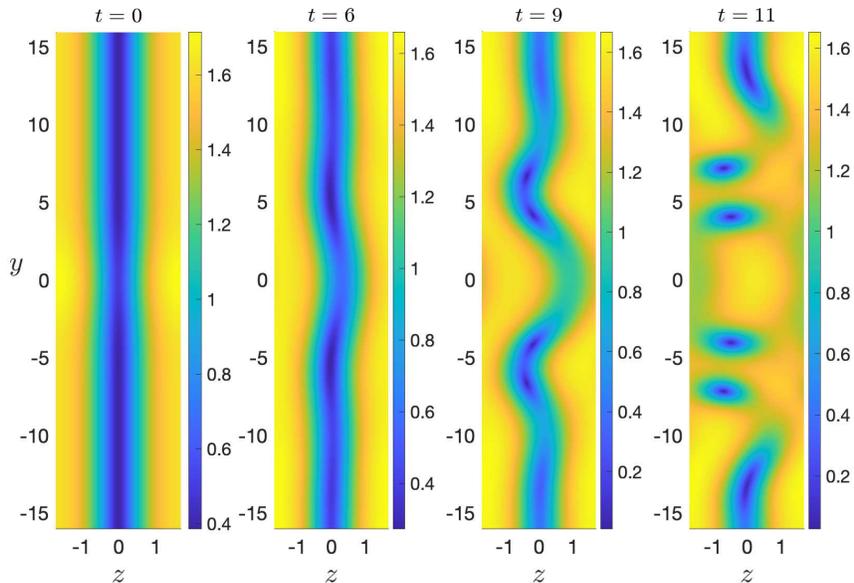}
\caption{Evolution of a perturbed traveling wave in elliptic ($s = 1$) NLS eq.~(\ref{NLS_phase_brought_down}). Plotted is $|\Phi(z,y,t)|$ for initial condition (\ref{NLS_ICs}) with parameters: $m = 0.75, b = 3, \epsilon = 1$ and period $1/k \approx 3.37.$}
\label{ellip_evolve}
\end{centering}
\end{figure}

\par The second typical instability pattern is displayed in Fig.~\ref{ellip_evolve} for the elliptic NLS equation. Here a ``snake" type instability  develops, where the function takes an oscillatory form in the transverse direction. Subsequently, filament dips shed from the snake profile. We point out the elliptic instability takes longer to develop than that of the hyperbolic case, in agreement with our stability analysis above. Finally, we note that these instability patterns found with the nontrivial phase solutions are of the same type as those identified in the simple phase solution \cite{CarSeg03}.

\section{Conclusion}
In this paper the Whitham theory for the (2+1)-dimensional elliptic defocusing/hyperbolic NLS equation (2dNLS) has been developed. These Whitham equations (\ref{eq:Wr-eq}) in hydrodynamic variables differ from the 1dNLS theory by introduction of $y$-derivative terms and nondiagonal terms (i.e.~the `$h_j$' terms); there are also two other dynamical equations (\ref{eq:q-eq}), (\ref{eq:p-eq}) and two constraints (\ref{eq:kq-eq}), (\ref{eq:px-eq}). We found that this system is unstable to transverse variations. This theory was compared with direct numerical calculations; very good agreement was obtained. This provides an independent check of our calculations. It also demonstrates that 2dNLS DSW propagation is different from the 1d NLS DSW propagation which does not have instability in the defocusing regime. 
\par This transverse linear instability of periodic traveling waves also suggests potential difficulties in observing 2d NLS DSWs. If wide transverse variation is allowed then this transverse instability could be a serious issue. In some experiments the transverse scale is circularly symmetric e.g.~\cite{HoeferEtAl06,WJF-NatPhys07}; in these cases the radial NLS equation \cite{rNLS} is a preferred model. Some phenomena are quasi-one dimensional e.g.~recent experiments of ``dam-break flow" of a photon fluid in fiber optics~\cite{TrilloXuEtAl17} or magnetic film experiments involving spin wave step pulse evolution into a DSW~\cite{JanSprenHoefWu17}. In these and other problems such as fiber optics where the transverse scale is limited, the one dimensional theory works well \cite{Ab2011}. There are also reductions corresponding to the problem of flow around an obstacle where stationary solutions involving quasi-one-dimensional type Whitham theory of dispersive shock waves in $x,y$ coordinates (one of which plays the role of effective time) are found, see~\cite{HoefElKam17} and references therein. On the other hand, often one needs a two- or three-dimensional theory in order to understand the underlying complex phenomena. In this respect our  2d NLS Whitham theory appears promising. There are also known interesting (2+1)-dimensional reductions of 2d NLS equation that have been studied, e.g.~the KP limit for small variations of a nonzero constant background, e.g.~\cite{KuzTur88, KuzRas95, PSK95, ChSch18}, which would lead us to the KP Whitham system of~\cite{ABW-KP, ABR}. 

\par It is possible that for defocusing elliptic and hyperbolic 2d NLS equations there may be few if any stable exact solutions. Nevertheless intermediate dynamics can exhibit rich and interesting structures; see e.g.~the numerics in~\cite{HyperNLS} for the hyperbolic case. It might be possible to investigate approximate intermediate solutions with the help of the 2d NLS Whitham system. The issues related to the presence of both focusing and defocusing directions in the hyperbolic NLS and an apparent vacuum point occurring there for the propagation with slope $q=\pm1$ (see Remark 3 in section 2) are also a matter for further investigation.
 
\par The 2dNLS Whitham theory suggests rich and novel phenomena. The detailed analytical and numerical study of these phenomena is a matter of future work. There are numerous interesting aspects to explore. There is considerable evidence that linear instabilities of e.g.~dark solitons of elliptic 2d NLS equation can develop into vortices and vortex-antivortex pairs, see e.g.~\cite{KuzRas95, PSK95, TheoEtAl03, ChSch18} and also our Fig.~\ref{ellip_evolve}. There are indications in these papers of some single branches of solutions connecting plane or ring dark solitons and vortices in the defocusing elliptic 2d NLS. The importance of vortices and related phenomena is well-known e.g.~in the physics of BECs~\cite{PitStrBEC}. Solutions describing vortices and related phenomena can be explored with the help of the (2+1)-dimensional Whitham theory. In any case, looking for solutions of the (2+1)-dimensional Whitham systems related to the 2d NLS as well as KP-type PDEs, are important problems for the future development of this theory.

\bigskip
{\bf\large Acknowledgments}  \\
This work was partially supported by NSF under grants DMS-1712793 and DMS-2005343.

\def\thesection{Appendix~\Alph{section}:}


\appendix

\section{Formulas for the elliptic solution}    \label{a:El}

Consider the elliptic leading order equation (\ref{eq:2.10}) for the density $\rho_0=|\Psi_0|^2$ (let $q=0$ for now, then $\rho_0=\mu_0$ and we are in 1d NLS situation, otherwise all the formulas below apply to $\mu_0=\rho_0/(1+sq^2)$),

\be
k^2(\rho_0')^2 = 2(\rho_0^3 - e_1\rho_0^2 + e_2\rho_0 - e_3).   \label{eq:A1}   
\ee

\ni Its derivative in $\theta$ is

\be
k^2\rho_0'' = 3\rho_0^2 - 2e_1\rho_0 + e_2.  \label{eq:A2}   
\ee

\ni Integrating eq.~(\ref{eq:A2}) over the period in $\theta$, one finds

\be
N \equiv \int_0^1\rho_0^2d\theta = \frac{2e_1Q-e_2}{3} = \frac{2}{3}(2\Omega Q - C_1).   \label{eq:A3}   
\ee

\ni Dividing eq.~(\ref{eq:A1}) by $\rho_0$ and then integrating over the period in $\theta$, one obtains

$$
\int_0^1\frac{k^2(\rho_0')^2+2e_3}{\rho_0}d\theta = k^2\int_0^1\frac{4C_0^2 + (\rho_0')^2}{\rho_0}d\theta = 2(N-e_1Q+e_2),
$$

\ni and, using eq.~(\ref{eq:A3}),

\be
\int_0^1\frac{k^2(\rho_0')^2+2e_3}{\rho_0}d\theta = k^2\int_0^1\frac{4C_0^2 + (\rho_0')^2}{\rho_0}d\theta = \frac{2(2e_2-e_1Q)}{3} = \frac{4(2C_1 - \Omega Q)}{3}.   \label{eq:A4}   
\ee

\ni Another formula used in the main text may be obtained e.g.~if one takes the combination of eqs.~$($\ref{eq:A1}$) - 2\rho_0\cdot($\ref{eq:A2}$)$, divides it by $\rho_0^2$ and then integrates over the period. The result is

\be
\int_0^1\frac{k^2(\rho_0')^2-2e_3}{\rho_0^2}d\theta = k^2\int_0^1\frac{(\rho_0')^2-4C_0^2}{\rho_0^2}d\theta = 4Q - 2e_1 = 4Q - 4\Omega.   \label{eq:A5}   
\ee



The leading order velocity $u_0$, see eq.~(\ref{eq:3.6}), averaged over the period in $\theta$ can be expressed in terms of the third complete elliptic integral:


\be 
\int_0^1u_0d\theta = \frac{V}{2} - \frac{C_0}{2}\int_0^1\frac{d\theta}{\mu_0} = \frac{V}{2} - \sigma\frac{\sqrt{\lambda_2\lambda_3}}{\sqrt{2\lambda_1}}\frac{\Pi(\gamma, m)}{K(m)},    \label{eq:A7}   
\ee
where $\Pi(\gamma, m)$ is the third complete elliptic integral and

$$
\gamma = -\frac{\lambda_2-\lambda_1}{\lambda_1},   \qquad   m = \frac{\lambda_2-\lambda_1}{\lambda_3-\lambda_1}.
$$
The third complete elliptic integral, cf.~e.g.~\cite{BF71}

\be
\Pi(\gamma,m) = \int_0^{K(m)}\frac{dz}{1-\gamma\;\text{sn}^2(z;m)},   \label{eq:Pi}
\ee
has the following derivatives with respect to its two arguments:

\be
\frac{\prt\Pi(\gamma,m)}{\prt m} = \frac{E - (1-m)\Pi}{2(1-m)(m-\gamma)},   \label{eq:mPi}
\ee

\be
\frac{\prt\Pi(\gamma,m)}{\prt \gamma} = \frac{(m-\gamma^2)\Pi - \gamma E - (m-\gamma)K}{2\gamma(1-\gamma)(m-\gamma)}.   \label{eq:gPi}
\ee

\par The other complete elliptic integrals have the following limiting expressions:

\be
m \to 0:  \qquad  K(m) = \frac{\pi}{2}\left(1 + \frac{m}{4} + \frac{9m^2}{64} + \dots\right),  \qquad E(m) = \frac{\pi}{2}\left(1 - \frac{m}{4} - \frac{3m^2}{64} + \dots\right);  \label{eq:A8}   
\ee

\be
m \to 1:  \qquad K(m) \approx \frac{1}{2}\ln\frac{16}{1-m},  \qquad  E(m) \approx 1 + \frac{(1-m)}{4}\left(\ln\frac{16}{1-m} - 1\right).  \label{eq:A9}   
\ee

\ni They satisfy the following differential equations in $m$:

\be
K'(m) = \frac{E-(1-m)K}{2m(1-m)},   \qquad  E'(m) = -\frac{K-E}{2m}.   \label{eq:A10}   
\ee





\section{Transformation to $D-D_x-D_y$ system}   \label{a:DDxDy}

Here, as an intermediate first step in obtaining the 2d NLS Whitham system in Riemann-type variables, we derive the eqs.~(\ref{eq:tk})--(\ref{eq:qp}) below. The following relations and notations are used later:

\be
V_y = s(qV+2p),  \qquad R_2 = V^2 + 4\Omega,  \qquad R_{2y} = (qV+2p)^2 + 4q^2\Omega,     \label{eq:RVy} 
\ee

\be
E_1 \equiv \left(V^2 + \frac{4\Omega}{3}\right)Q - 2VC_0 + \frac{4C_1}{3} = V(P - C_0) + 2C_1 + N = B+2N,  \qquad 3N = 4\Omega Q - 2C_1,    \label{eq:E1def}
\ee

{\small\be
H = gE_1 + 4sp(qP + pQ),   \quad  G \equiv 4(V+qV_y)C_1 - (R_2 + sR_{2y})C_0 = gG_1 + 4sp[q(2C_1 - VC_0) - pC_0],  \quad  G_1 \equiv 4VC_1 - R_2C_0.   \label{eq:HG1}  
\ee}

$$
P = VQ - C_0,   \qquad P_y = (qV+2p)Q - qC_0,   \qquad  H = VP + V_yP_y - (V+qV_y)C_0 + g(2C_1+N).    
$$
It will be convenient here to introduce and use the operators defined in eq.~(\ref{eq:Ds}),

$$
D = \prt_t + V\prt_x + V_y\prt_y,   \qquad  D_x = \prt_x + sq\prt_y,   \qquad  D_y = \prt_y - q\prt_x,       
$$
rather than partial derivatives $\prt_t$, $\prt_x$ and $\prt_y$. By combining the Whitham equations (\ref{eq:(k)})--(\ref{eq:E}) we will express them in terms of operators $D$, $D_x$ and $D_y$, and will eventually bring them to the following form called ``$D-D_x-D_y$" system:

\be
g\left(\frac{Dk}{k} + D_xV\right) + sq[Dq + VD_xq + 2D_xp] = 0,   \label{eq:tk}   
\ee

$$
g\left[4\Omega DQ - 2QD\Omega - 2DC_1 + 3C_0DV - (4\Omega Q-8C_1-3VC_0)\frac{Dk}{k} + \frac{3}{2}C_0D_xR_2\right] +
$$

\be
+ 3sqC_0[2VDq + R_2D_xq + 2(Dp + VD_xp)] = 0,   \label{eq:KR2}
\ee

\be
g\left(DQ - Q\frac{Dk}{k} - D_xC_0\right) + sq[QDq - 3C_0D_xq] + s[PD_yq + 2QD_yp] = 0,   \label{eq:ttQ}
\ee

$$
g\left(QDV - DC_0 + 2C_0\frac{Dk}{k} + 2D_xC_1\right) + 
$$

\be
+ sq[PDq + 2QDp + 8C_1D_xq] + s[(2C_1-N-VC_0)D_yq - 2C_0D_yp] = 0,   \label{eq:ttP2}
\ee

\be
Dq + gD_yV + sq(VD_yq + 2D_yp) = 0,   \label{eq:tq}
\ee


\be
PDq + 2QDp + (2C_1-VC_0)D_xq - 2C_0D_xp + sq(4\Omega Q-2C_1)D_yq + g\left(D_yN-N\frac{D_yk}{k}\right) = 0,   \label{eq:qp}
\ee
and the constraint obtained from equation (\ref{eq:kq})

\be
D_xq = g\frac{D_yk}{k} + sqD_yq   \label{eq:tkq}
\ee
and also the constraint given by eq.~(\ref{eq:pxl}). \\

\par The rest of this appendix is the derivation of the $D-D_x-D_y$ system above. In order to get equations (\ref{eq:tk})--(\ref{eq:qp}) we first rewrite the eqs.~(\ref{eq:Q}), (\ref{eq:P}), (\ref{eq:Py}) and (\ref{eq:E}) in the form 

\be
\prt_t(gQ) + \prt_x[g(VQ-C_0)] + \prt_y[g(V_yQ-sqC_0)] = 0,   \label{eq:Qa}
\ee

\be
\prt_t(gP) + \prt_x[g(V(P-C_0) + 2C_1 - N) + g^2N] + \prt_y[g(V_y(P-C_0) + 2spC_0 + sq(2C_1-N))] = 0,  \label{eq:Pa}
\ee

\be
\prt_t(gP_y) + \prt_x[g(V(P_y-qC_0) - 2pC_0 + q(2C_1 - N))] + \prt_y[g^2N + g(V_y(P_y-qC_0) + sq^2(2C_1-N))] = 0,  \label{eq:Pya}
\ee

\be
\prt_t(gH) + \prt_x[g(VH + G - 4sqpN)] + \prt_y[g(V_yH + sqG + 4spN)] = 0,     \label{eq:Ha}    
\ee
see the relations/notations in the beginning of the section. 
\par Using the operators from eq.~(\ref{eq:Ds}), we rewrite eq.~(\ref{eq:Qa}) as

$$
D(gQ) + gQ(\prt_xV+\prt_yV_y) - D_x(gC_0) - sgC_0\prt_yq = 0.
$$
Then we use the relations

\be
\prt_xV+\prt_yV_y = D_xV + s(V\prt_yq + 2\prt_yp),   \label{eq:divV}
\ee

\be
g\prt_x = D_x - sqD_y,   \qquad  g\prt_y = D_y + qD_x,   \label{eq:gDs}
\ee
to bring it to the form

\be
D(gQ) + gQD_xV - D_x(gC_0) + sP(D_yq + qD_xq) + 2sQ(D_yp + qD_xp) = 0.   \label{eq:QaD}
\ee
Quite similarly, we bring eq.~(\ref{eq:Pa}) to the form


$$
D(gP) + g(P-C_0)D_xV - VD_x(gC_0) + 2D_x(gC_1) - sqD_y(gN) + 
$$

\be
+ sq[V(P-C_0) + 2C_1 + N]D_xq + s[V(P-C_0) + 2C_1 - (1+2sq^2)N]D_yq + 2sP[qD_xp+D_yp] = 0.   \label{eq:PaD}
\ee
Also similarly, eq.~(\ref{eq:Ha}) is rewritten as


$$
D(gH) + gHD_xV + D_x(gG) + 4spD_y(gN) + s[q(VH+G) - 4pN]D_xq + s[VH+G + 4sqpN]D_yq + 
$$

\be
+ 2s[qHD_xp + (H + 2gN)D_yp] = 0.   \label{eq:HaD}
\ee
Then we combine in a certain way the equations obtained so far. Taking the combination of eqs.~$(\ref{eq:Pya}) - q(\ref{eq:Pa}) - 2p(\ref{eq:Qa})$, we get an equation with only $q$ and $p$ time derivatives. We use eqs.~(\ref{eq:Ds}), (\ref{eq:tkq}) and $3N=4\Omega Q-2C_1$ to rewrite the combination as eq.~(\ref{eq:qp}). Combination of eqs.~$(\ref{eq:PaD})-V(\ref{eq:QaD})$ yields


$$
gQDV - D(gC_0) - 2gC_0D_xV + 2D_x(gC_1) - sqD_y(gN) + 
$$

\be
+ sq(2C_1+N-VC_0)D_xq + s[2C_1-(1+2sq^2)N-VC_0]D_yq - 2sC_0[qD_xp + D_yp] = 0.   \label{eq:P2D}
\ee
Combination of equations $(\ref{eq:HaD}) - 4sqp(\ref{eq:PaD}) - 4sp^2(\ref{eq:QaD})-4sgp(\ref{eq:qp})$ leads, after using eq.~(\ref{eq:HG1}) and some tedious algebra, to the equation 







$$
D(gE_1) + gE_1D_xV + D_x(gG_1) + s[2qE_1Dq + q(VE_1+3G_1)D_xq + (VE_1+G_1)D_yq] + 
$$

\be
+ 2s[2qPDp + q(E_1+4C_1-2VC_0)D_xp + (E_1+2N)D_yp] = 0.   \label{eq:H1a}
\ee

Eq.~(\ref{eq:tq}) is obtained from eq.~(\ref{eq:(q)}) applying (\ref{eq:Ds}). Eq.~(\ref{eq:(k)}) is transformed, using eqs.~(\ref{eq:Ds}), (\ref{eq:gDs}), (\ref{eq:tq}) and (\ref{eq:kq}), to the form of eq.~(\ref{eq:tk}). Combining equations $(\ref{eq:QaD})-Q(\ref{eq:tk})$ yields eq.~(\ref{eq:ttQ}), to use further instead of eq.~(\ref{eq:QaD}). The combination $(\ref{eq:P2D})+2C_0(\ref{eq:tk})+sq(\ref{eq:qp})$ reduces to eq.~(\ref{eq:ttP2}). It remains to derive eq.~(\ref{eq:KR2}). 




\par From now on we use eqs.~(\ref{eq:tk}), (\ref{eq:tq}), (\ref{eq:qp}), (\ref{eq:ttQ}), (\ref{eq:ttP2}) and (\ref{eq:H1a}). Combining eqs.~$(\ref{eq:H1a})-(E_1+4C_1)(\ref{eq:tk})$ gives 



{\small$$
D(gE_1) - g(E_1+4C_1)\frac{Dk}{k} + 4VD_x(gC_1) - R_2D_x(gC_0) - gC_0D_xR_2 +  
$$

\be
+ s[q(E_1-4C_1)Dq + q(2G_1-R_2C_0)D_xq + (VE_1+G_1)D_yq] + 2s[2qPDp - 2qVC_0D_xp + (E_1+2N)D_yp] = 0.   \label{eq:tH1}
\ee}
Then taking the difference $(\ref{eq:tH1})-R_2(\ref{eq:ttQ})$ gives

{\small\be
D(gE_1) - g(E_1+4C_1)\frac{Dk}{k} - R_2\left(D(gQ) - gQ\frac{Dk}{k}\right) + 4VD_x(gC_1) - gC_0D_xR_2 +     \label{eq:ttH1}
\ee

$$
+ s[q(E_1-4C_1+R_2Q)Dq + 2qG_1D_xq + (VE_1+G_1-R_2P)D_yq] + 2s[2q(PDp - VC_0D_xp) + (E_1+2N-R_2Q)D_yp] = 0.   
$$}
Then the difference $(\ref{eq:ttH1})-2V(\ref{eq:ttP2})$ gives an equation equivalent to




$$
g\left[DE_1 - (E_1+4C_1)\frac{Dk}{k} - R_2DQ + gQ\frac{Dk}{k} - 2V\left(QDV - DC_0 + 2C_0\frac{Dk}{k}\right) - C_0D_xR_2\right] +  
$$

$$
+ s[q(3E_1-4C_1-R_2Q-2V(VQ-C_0))Dq - 2qR_2C_0D_xq + V(E_1-R_2Q+2N+2VC_0)D_yq] + 
$$

\be
+ 2s[-2qC_0(Dp + VD_xp) + (E_1+2N-R_2Q+2VC_0)D_yp] = 0.   \label{eq:ttKR2}
\ee
After recalling the definitions of $R_2$, $E_1$ and $N$ from eqs.~(\ref{eq:RVy}), (\ref{eq:E1def}), this simplifies to become eq.~(\ref{eq:KR2}). 


\par Finally this yields the system consisting of eqs.~(\ref{eq:tk})--(\ref{eq:qp}).  

\section{Transformation to Riemann variables}    \label{a:Rv}

We introduce the quantities $W_n, n=1,\dots,4$ given by eq.~(\ref{eq:W1234}). Then eq.~(\ref{eq:tk}) can be written as
\be
g\left(4\frac{Dk}{k} + D_xp_1 + W_1\right) = 0.   \label{eq:tk'}   
\ee
Eq.~(\ref{eq:KR2}) simplifies due to the identity

\be
\frac{4\Omega DQ-2QD\Omega-2DC_1-(4\Omega Q-8C_1)Dk/k}{3C_0} + DV + V\frac{Dk}{k} = \sum_jr_j\frac{\prt_jk}{k}Dr_j   \label{eq:Id2*}
\ee 
and takes the form

\be
\frac{3C_0}{4}g\left(4\sum_jr_j\frac{\prt_jk}{k}Dr_j + \frac{D_xp_2}{2} + W_2\right) = 0.   \label{eq:KR2'}
\ee
Then, taking the combination $8V/(3C_0)\cdot(\ref{eq:KR2}) - 16\cdot(\ref{eq:ttQ}) + 4(R_2-2V^2)\cdot(\ref{eq:tk})$ yields

$$
g\left[8V\left(\frac{4\Omega DQ-2QD\Omega-2DC_1-(4\Omega Q-8C_1)Dk/k}{3C_0} + DV + V\frac{Dk}{k}\right) -  \right.
$$

\be
\left. - 16\left(DQ-Q\frac{Dk}{k}\right) + 4(R_2-2V^2)\frac{Dk}{k} + \frac{D_xp_3}{3} + W_3\right] = 0.   \label{eq:KR3}
\ee
or, after transformations,

\be
g\left(4\sum_jr_j^2\frac{\prt_jk}{k}Dr_j + \frac{D_xp_3}{3} + W_3\right) = 0.   \label{eq:KR3'}
\ee
Finally, taking the combination $4(R_2+2V^2)/(3C_0)\cdot(\ref{eq:KR2}) + 16\cdot(\ref{eq:ttP2}) - 48V\cdot(\ref{eq:ttQ}) + 8(6C_0+VR_2 - 2V^3)\cdot(\ref{eq:tk})$ yields

{\small$$
g\left[4(R_2+2V^2)\left(\frac{4\Omega DQ-2QD\Omega-2DC_1-(4\Omega Q-8C_1)Dk/k}{3C_0} + DV + V\frac{Dk}{k}\right) + \right.
$$}

{\small\be
\left. + 16\left(QDV - DC_0 + 2C_0\frac{Dk}{k}\right) - 48V\left(DQ-Q\frac{Dk}{k}\right) +  8(6C_0+VR_2 - 2V^3)\frac{Dk}{k} + \frac{D_xp_4}{4} + W_4 \right] = 0,  \label{eq:KR4}
\ee}
or, after transformations,

\be
g\left(4\sum_jr_j^3\frac{\prt_jk}{k}Dr_j + \frac{D_xp_4}{4} + W_4\right) = 0.   \label{eq:KR4'}
\ee

\par Thus, the four of the Whitham equations, (\ref{eq:tk}), (\ref{eq:KR2}), (\ref{eq:ttQ}) and (\ref{eq:ttP2}), are transformed to eqs.~(\ref{eq:DWr}) where the log-derivatives $\frac{\prt_jk}{k} \equiv \frac{\prt\ln k}{\prt r_j}$ are given by eq.~(\ref{eq:4.6}). Eqs.~(\ref{eq:DWr}) can be written in matrix form as






$$
\Delta_{jl}\left(4\frac{\prt_lk}{k}Dr_l + D_xr_l\right) + W_j = 0,
$$

\ni where $\Delta$ is the Vandermonde matrix,

{\small $$
\Delta = \left( \begin{array}{cccc} 1 & 1 & 1 & 1 \\ r_1 & r_2 & r_3 & r_4 \\  r_1^2 & r_2^2 & r_3^2 & r_4^2 \\  r_1^3 & r_2^3 & r_3^3 & r_4^3 \end{array} \right),
$$}

\ni and the vector $W=(W_1, W_2, W_3, W_4)^T$ has components given in eq.~(\ref{eq:W1234}). Inverting the Vandermonde matrix, we obtain the four equations of the form 

\be
4\frac{\prt_jk}{k}Dr_j + D_xr_j + \tilde h_j(r_1,r_2,r_3,r_4) = 0,   \quad j=1,2,3,4,  \label{eq:Wh}   
\ee
each containing $D$- and $D_x$-derivatives of only one $r_j$-variable. The terms $\tilde h_j$ would be absent in the 1d NLS case. So this last transformation would diagonalize the Whitham equations for 1d NLS bringing them to the well-known form, see e.g.~\cite{HoeferEtAl06, ElKr95}. Here 





\be
\tilde h_j = \frac{\Delta_{ilm}}{|\Delta|}(-r_ir_lr_mW_1 - (r_ir_l +r_lr_m+r_mr_i)W_2 - (r_i+r_l+r_m)W_3 + W_4),   \label{eq:hjB}   
\ee
where $j\neq i, j\neq l, j\neq m$, $i<l<m$ and

$$
|\Delta| = \det\Delta = \prod_{j>l}^4(r_j-r_l),  \qquad  \Delta_{jlm} = \left| \begin{array}{ccc} 1 & 1 & 1 \\ r_j & r_l & r_m \\  r_j^2 & r_l^2 & r_m^2 \end{array} \right| = (r_j-r_l)(r_l-r_m)(r_m-r_j),
$$
are the determinant and the corresponding minors of the Vandermonde matrix. Thus, dividing eqs.~(\ref{eq:Wh}) by $4\prt_jk/k$ and using eqs.~(\ref{eq:Ds}), we get the Whitham eqs.~(\ref{eq:Wr}) in terms of $r$-variables.
\par Next we transform eqs.~(\ref{eq:pxl}) and (\ref{eq:qp}). We use the identity

\be
\frac{D_y\lambda_1}{\lambda_{21}\lambda_{31}} - \frac{D_y\lambda_2}{\lambda_{21}\lambda_{32}} + \frac{D_y\lambda_3}{\lambda_{32}\lambda_{31}} = -8\sum_j\frac{D_yr_j}{r_{ji}r_{jl}r_{jm}},    \label{eq:Id1}
\ee
which is derived from eq.~(\ref{eq:4.3}) and

\be
C_0\left( \frac{D_y\lambda_1}{\lambda_1\lambda_{21}\lambda_{31}} - \frac{D_y\lambda_2}{\lambda_2\lambda_{21}\lambda_{32}} + \frac{D_y\lambda_3}{\lambda_3\lambda_{32}\lambda_{31}} \right) = -8\sum_j\frac{(r_j-V)D_yr_j}{r_{ji}r_{jl}r_{jm}},    \label{eq:Id2}
\ee
which is a consequence of eqs.~(\ref{eq:4.3}) and (\ref{eq:C0r}), as well as eqs.~(\ref{eq:r3id}) and (\ref{eq:Q-j}); this brings eq.~(\ref{eq:pxl}) to the form of eq.~(\ref{eq:px-eq}). Adding eq.~(\ref{eq:qp}) and eq.~(\ref{eq:tq}) multiplied by $C_0$, and using again that $3N=4\Omega Q-2C_1$ and eq.~(\ref{eq:tkq}), we obtain

$$
Q\left(2Dp + VDq + 4sq\Omega D_yq + \frac{4}{3}gD_y\Omega\right) + gC_0\left(D_yV - V\frac{D_yk}{k} - 2\prt_xp\right) +
$$

\be
+ \frac{g}{3}\left[ 4\Omega\left(D_yQ - Q\frac{D_yk}{k}\right) - 2D_yC_1 + 8C_1\frac{D_yk}{k} \right] = 0.   \label{eq:qp*}
\ee
Then, after applying eqs.~(\ref{eq:px-eq}) and (\ref{eq:Id2*}), eq.~(\ref{eq:qp*}) simplifies to become eq.~(\ref{eq:qp-simple}).

\par Next we simplify the $h_j$-terms. Using the relations

$$
r_i+r_l+r_m = p_1-r_j = 4V-r_j,   \qquad  r_ir_l+r_lr_m+r_mr_i = r_j^2 - p_1r_j + \frac{p_1^2-p_2}{2} = r_j^2 - 4Vr_j + 2(4V^2-R_2),
$$

$$
r_ir_lr_m = -r_j^3 + p_1r_j^2 - \frac{p_1^2-p_2}{2}r_j + \frac{2p_3-3p_1p_2+p_1^3}{6} = -r_j^3 + 4Vr_j^2 - 2(4V^2-R_2)r_j - 4(4C_0+VR_2-2V^3),
$$

\be
r_{ji}r_{jl}r_{jm} = 4r_j^3 - 3p_1r_j^2 + (p_1^2-p_2)r_j - \frac{2p_3-3p_1p_2+p_1^3}{6} = 4[r_j^3 - 3Vr_j^2 + (4V^2-R_2)r_j + 4C_0+VR_2-2V^3],   \label{eq:r3id}
\ee
we can rewrite $h_j$-terms as follows: 

$$
h_j = \frac{1}{4r_{ji}r_{jl}r_{jm}\prt_jk/k}\left[ -r_j^3W_1 + r_j^2(4VW_1-W_2) + r_j((R_2-6V^2)W_1+2VW_2-\tilde W_3) - \right.
$$

$$
\left. -\tilde W_4 + V\tilde W_3 + (R_2-2V^2)W_2 - (4C_0+2VR_2-4V^3)W_1 \right],  
$$

\be
g\tilde W_3 = 16s[q(QDq - 3C_0D_xq) + PD_yq + 2QD_yp],   \label{eq:tW3}
\ee

\be
g\tilde W_4 = 16s[q(PDq + 2QDp + 8C_1D_xq) + (2C_1-N-VC_0)D_yq - 2C_0D_yp],   \label{eq:tW4}
\ee
which implies

$$
h_j = \frac{1}{4r_{ji}r_{jl}r_{jm}\prt_jk/k}\left[ -W_1r_{ji}r_{jl}r_{jm}/4 + r_j^2(VW_1-W_2) + r_j(2V(W_2-VW_1)-\tilde W_3) \right.
$$

$$
\left. - \tilde W_4 + V\tilde W_3 + (R_2-2V^2)(W_2 - VW_1) \right].
$$
Then we use the formula (which can be derived e.g.~from eqs.~(\ref{eq:QrE}) and (\ref{eq:4.6}))

\be
Q = \frac{1}{4}\left(r_{ji}r_{jl}r_{jm}\frac{\prt_jk}{k} - r_j^2 + 2Vr_j + R_2-2V^2\right),  \qquad  j=1,\dots, 4, \quad j\neq i\neq l\neq m\neq j,   \label{eq:Q-j}
\ee
and eqs.~(\ref{eq:r3id}) to transform $h_j$ to the expression in eq.~(\ref{eq:hj1}). Finally, we use the algebraic identity

$$
(R_2-V^2)(r_j^2 - 2Vr_j + 2V^2-R_2) - 12C_0(r_j-V) + 32C_1 = 
$$

\be
= (r_j-V)[r_j^3 - 3Vr_j^2 + (4V^2-R_2)r_j + 4C_0 + VR_2 - 2V^3],   \label{eq:r4id}
\ee
which can be verified with the expressions (\ref{eq:Ve1r})--(\ref{eq:e2r}), to bring eq.~(\ref{eq:hj1}) to the form of eq.~(\ref{eq:hj-eq}), i.e.~without denominator.

\end{document}